\documentclass[preprint,12pt]{article}

\usepackage{soul}
\usepackage{url}
\usepackage[hidelinks]{hyperref}
\usepackage[utf8]{inputenc}
\usepackage[small]{caption}
\usepackage{graphicx}
\usepackage{amsmath}
\usepackage{amsthm}
\usepackage{booktabs}
\usepackage{algorithm}
\usepackage{algorithmic}
\usepackage[switch]{lineno}
\usepackage[a4paper, total={6in, 8in}]{geometry}

\usepackage{caption}
\usepackage{subcaption}
\usepackage[numbers]{natbib} 
\usepackage{cleveref}
\usepackage{graphicx} 
\usepackage{booktabs}
\usepackage{amsthm}
\usepackage{amsfonts}
\newtheorem{definition}{Definition}
\usepackage{xcolor}
\usepackage{authblk}
\title{Downside Risk-Aware Equilibria for Strategic Decision-Making
\footnotetext{\textbf{Presented at ECAI 2024: Workshop on AI in Finance}}
}
\date{}

\author[1]{Oliver Slumbers\thanks{Work done while at JP Morgan AI Research, London, United Kingdom}}
\author[2]{Benjamin Patrick Evans\thanks{Corresponding author: benjamin.x.evans@jpmorgan.com}}
\author[3]{Sumitra Ganesh}
\author[2]{Leo Ardon}

\affil[1]{University College London, United Kingdom}
\affil[2]{JPMorgan AI Research, London, United Kingdom}
\affil[3]{JPMorgan AI Research, New York, USA}

\begin{document}

\maketitle

%% Abstract
\begin{abstract}
Game theory has traditionally had a relatively limited view of risk based on how a player's \textit{expected} reward is impacted by the uncertainty of the actions of other players.
Recently, a new game-theoretic approach provides a more holistic view of risk also considering the reward-variance. However, these variance-based approaches measure variance of the reward on both the upside and downside. In many domains, such as finance, downside risk \textit{only} is of key importance, as this represents the potential losses associated with a decision. In contrast, large upside "risk" (e.g. profits) are not an issue.  To address this restrictive view of risk, we propose a novel solution concept, downside risk aware equilibria (DRAE) based on lower partial moments.  DRAE restricts downside risk, while  placing no restrictions on upside risk, and additionally, models higher-order risk preferences. We demonstrate the applicability of DRAE on several games, successfully finding equilibria which balance downside risk with expected reward, and prove the existence and optimality of this equilibria. 
\end{abstract}

\newpage
\section{Introduction}
Traditional game-theoretic approaches to risk are generally limited by their focus on expected reward \cite{bielefeld1988reexamination, mckelvey1995quantal}, in that they can often undervalue the impact of large costs to a risk-averse practitioner. \citet{slumbers2023game}
%took inspiration from financial decision-making, specifically Modern Portfolio Theory (MPT) \cite{10.2307/2975974}, to design a game-theoretic equilibrium concept, called
proposed Risk Aware Equilibria (RAE), that additionally considered reward variance (caused by the strategy of opposing players) as a risk metric for players. However, 
%in much the same way that MPT can be critiqued, RAE shares similar problems by 
%RAE only consideres the first two moments of the reward distribution. 
we stress two problems of such mean-variance approaches for risk-aware decision making, and a third limitation specific to the game-theoretic setting of \cite{slumbers2023game}: 1) by controlling the variance of the distribution, one controls both downside \textit{and} upside variance. In the large majority of scenarios, there is no incentive to limit upside variance (e.g. large profits in a financial setting). 2) mean-variance approaches assume that players only care about the first two moments of the reward distribution, when in reality, risk preferences are shown to be widely heterogeneous and higher-order \cite{egan2021drives}. 
%The practitioner should have the flexibility to optimise over higher-order risk preferences.
3) RAE has only considered endogenous risk due to the strategy of the opposing players, and not exogenous risk in the environment itself, for example, through risky states.

In this work, we aim to improve the models of and solution concepts for risk-aware decision-making in game-theory by addressing the limitations of the existing mean-variance approaches. Specifically, we propose an equilibria concept, Downside Risk-Aware Equilibria (DRAE). DRAE is a quadratic programming solvable equilibrium that considers (higher-order) lower \textit{partial} moments \cite{nawrocki1991optimal} of the reward distribution. We prove that DRAE corresponds to the equilibrium solution with the lowest possible downside risk given the provided risk-preferences, while still sharing the desirable theoretic properties of both the Nash equilibrium and RAE.

%and the risk inherent to the environment (e.g. asset price fluctuations). The equilibria concept allows for player risk preferences to be characterised by more than just mean-variance, meaning higher moments of the distribution must be considered.
%In designing the risk component of our equilibrium concept we have two general desiderata: 1) improved flexibility to model diverse risk preferences through incorporating higher order moments, and 2) to not actively constrain the freedom of the upside of a reward.To achieve this, we 
%build upon results from the finance literature \cite{cumova2011symmetric, estrada2004mean, nawrocki1992characteristics} to

Across various environments, we validate that DRAE is able to successfully balance expected reward and downside risk. In all of our environments, DRAE has significantly less downside risk in comparison to both the Nash Equilibrium and RAE (which in some instances can have near unbounded downside risk), while maintaining the expected reward. We further demonstrate the benefit and flexibility of DRAE by modelling higher order risk preferences, which isn't possible with existing solution concepts. Finally, we show that DRAE is able to account for the presence of exogenous risk from the environment, successfully managing both \textit{endogenous} and \textit{exogenous} risk.

\section{Background and Related Work}

\citet{harsanyi1988general} introduced risk-dominant Nash equilibria (NE) \cite{nash1951non}, which, when the strategies of other players are unknown, leads to the NE with the lowest losses if players deviate from the NE solution. However, if none of the NE are robust to risk initially, then the risk-dominant strategy among NEs will also not be. Trembling-hand Perfect Equilibrium (THPE) \cite{bielefeld1988reexamination} models risk by having players 'tremble', with all actions receiving positive probability in a mixed-strategy. However, iterated deletion of weakly dominated strategies can lead to the THPE strategy being removed from consideration. \citet{mckelvey1995quantal} introduce Quantal-Response equilibrium (QRE), accounting for potential errors in strategy selection based on the magnitude of the expected reward (ER). Whilst risk aversion is not the original intention of QRE, modelling for errors has similar risk interpretations to that of THPE and has been investigated in the context of MARL in 
\cite{mazumdar2024a}. THPE and QRE both utilise ER as their risk measure which is insensitive to low probability, high-value deviations as long as they are offset by high probability mean values \cite{royset2022risk}. \citet{slumbers2023game, yekkehkhany2020risk, xu2024balancing} attempt to remedy the problems of only considering ER through the introduction of reward variance as a risk measure. \cite{slumbers2023game} considers the reward variance in line with that of THPE and QRE, and introduces RAE which considers how reward varies in response to the actions of the other players. The approach aims to find strategies that have high ER whilst being robust to any action that the opponent may take (no opponent action will cause reward to largely deviate from ER). Likewise, \cite{xu2024balancing} utilises variance as a risk measure in a modified version of Counterfactual Regret Minimization. \cite{yekkehkhany2020risk} instead considers the problem from a probabilistic perspective, where the utilities are not deterministic and instead drawn from a probability distribution. This is a different form of risk and not that which is considered by THPE, QRE, NE or RAE.

\subsection{Beyond mean-variance}

One common area balancing risk and return is portfolio theory. Mean-variance portfolio theory \cite{10.2307/2975974} is a framework for designing investment portfolios to maximise return for a given risk (in the original case, variance of the return). However, there are two key limitations of using variance based risk metrics, as discussed earlier. Using semi-variance (only considering rewards below the mean) was proposed as an alternative \cite{10.2307/2975974}, however not utilised due to computational difficulties. Approaches to manage the computational difficulties of semi-variance, specifically focusing on the estimation of the co-semivariance matrix required for a quadratic programming solution, have since been proposed \cite{estrada2004mean, ferrari2024smoothed, markowitz2020avoiding}, all functioning well under slightly different conditions on the rewards.

Generalisations of the semi-variance have then been made 
%Furthermore, whilst less-explored, efforts have been made to explore further semi-moments other than semi-variance in the portfolio optimisation setting 
based on lower partial moments (LPM). LPM are measures over functions that only consider values that falls below a given threshold $\tau$. By selecting the degree $d$ of the LPM, one can both model higher-order moments of a function (e.g. skewness, kurtosis) and model higher-order risk attitudes, for example, attitudes such as prudence and temperance are associated with the third and fourth moments \cite{niguez2015higher}. Importantly, LPM with a suitable threshold, do not actively constrain the upside risk, for example \textit{not} penalising negatively skewed (right-tailed) return distributions, which a variance measure is ill-equipped to do. Similarly to the semi-variance case, computation of the co-LPM matrix used for quadratic programming optimisation is difficult, with some work focusing on these computational issues \cite{mondal2022convexity, nawrocki1991optimal, nawrocki1992characteristics}.

Thus far, LPM has not been used as a tool for measuring risk in game-theoretic settings. 
%As has been found in the finance literature, incorporating LPM as a risk measure generally is non-trivial due to the properties of the LPM matrix not being as easy to work with as the co-variance matrix. 
In the next section, we introduce our approach for integrating LPM for risk-aware decision-making in game-theory.

%\subsection{Motivating Example}

%\begin{figure}
%    \centering
    %\includegraphics[width=.5\columnwidth]{extended_abstract/plots/normal.pdf}
    %\caption{Normally distributed returns, with a high and low risk setting}
    \label{figNormal}
%\end{figure}

%\begin{figure}
 %   \centering
    %\includegraphics[width=.5\columnwidth]{extended_abstract/plots/skew.pdf}
    %\caption{Skewed returns, with equal mean and variance, but an upside and downside risk setting.}
    %\label{figSkew}
%\end{figure}

%\cref{figNormal} visualises a high and low risk normal distribution setting. The high risk line has higher expected utility, so would be chosen in most game theoretic settings, however, the low risk line has lower variance, so minimises the potential risk and may be preferred for certain investment strategies. Such a setting is well captured by the RAE introduced in \cite{slumbers2023game}. However, consider also that returns may be non-normally distributed, e.g. \cref{figSkew} now considers a skew, and we may be more interested in minimising the downside risk. Both settings in \cref{figSkew} have the same mean and standard deviation, but vary in the resulting skew -- with one having higher upside "risk" and one higher downside risk. If we wanted to select the option that minimises the potential financial downside (with higher upside "risk"), this would not be possible under the existing frameworks as the two appear indistinguishable from the first two moments alone. Such considerations motivate this work: How can we model downside risk when selecting actions in a game-theoretic setting?

%This can be achieved by utilising semi-variance measures such as \textbf{Lower partial moments}, as detailed in the following section.

%%%%%%%%%%%%%%%%%%%%%%%%%%%%%%%%%%%%%%%%%%%%%%%%%%%%%%%%%%%%%%%%%%%%%%%%

\section{Proposed Approach}
We propose a novel downside risk-aware equilibrium (DRAE) solution concept, leveraging both the LPM advances in risk management and the equilibrium approach of \cite{slumbers2023game}. DRAE significantly improves upon RAE \cite{slumbers2023game}, addressing its primary deficiencies. Importantly, improving upon the risk metric used in RAE, by replacing the variance with an LPM metric to only consider downside risk with a more flexible model of higher-order preferences. Additionally, we account for risk exogenous to the environment (e.g. asset price randomness), rather than just considering the risk induced by opponent strategy selection. 

In this section, we introduce the formulation of DRAE for normal-form games (NFG). A NFG is the standard representation of strategic interaction in game-theory, which we extend to a state-based system to additionally account for environment risk. A finite state-based $n$-person NFG can be represented a tuple $(N, A, S, r)$
 where $N$ is a finite set of $n$ players, $A = A^1 \times ... \times A^n$ is the joint action space, with $A^i$ being the actions available to player $i$, $S$ is a finite set of $s$ states which occur with probability $q(s)$ such that $\sum_{s \in S} q(s) = 1$, and $r = (r^1,...,r^n)$ where $r^i: A \times S \rightarrow \mathbb{R}$ is the real-valued reward function for each player $i$. Here, rewards are the values that come directly from the environment, such as asset returns. Utility, referenced later, more generically refers to the evaluation of these rewards by combining ER with a risk measure. A player plays a mixed-strategy, $\boldsymbol{\sigma}^i \in \Delta_{A^i}$ which is a probability distribution over their action space $A^i$.

 \subsection{DRAE}
The objective of DRAE is to provide an equilibrium solution that maximises the expected reward for a player whilst minimising their potential \emph{downside} risk from both the environment and the opponent, maximising their expected utility. 

For simplicity, and without loss of generality, we formulate DRAE based on symmetric two-player games, such that two players share an action set $A^1 = A^2 = A$, and a reward function $r^1 = r^2 = r$. We also assume that states $s$ are global and always the same between the players. The reward of action $a_m \in A$ against action $a_n \in A$ in state $s \in S$ is defined as $r(a_m, a_n, s)$ and the full reward matrix as $\mathbf{M}(s)$, where the entry $\mathbf{M}_{m,n}(s)$ refers to $r(a_m, a_n, s)$ and $\mathbf{M}_m(s)$ refers to $r(a_m, a, s) \; \forall a \in A$, i.e., the vector of rewards that action $a_m$ receives against all other actions in state $s$. The the expected reward of the mixed strategy for player 1 $\boldsymbol{\sigma}$ versus the mixed strategy for player 2 $\boldsymbol{\varsigma}$ is define as:

 \begin{align}
    \mathrm{ER}(\boldsymbol{\sigma}, \boldsymbol{\varsigma}, \boldsymbol{M}, S) &= \sum_{s\in S} q(s) \sum_{a_i \in A} \sum_{a_j \in A} \boldsymbol{\sigma}(a_i) \boldsymbol{\varsigma}(a_j) r(a_i, a_j, s) \\
    &= \boldsymbol{\sigma}^T \cdot \boldsymbol{M} \cdot \boldsymbol{\varsigma}
\end{align}

\textbf{Risk}
Risk is based on the LPM of each action $a \in A$. In order to use a LPM measure, a threshold value $\tau$ which quantifies which rewards are 'lower' and those that are 'upper' must be defined. The threshold $\tau$ is environment dependent, and examples include historical means of expected reward, expected reward under a random strategy, or risk-free return rates. The LPM of an action $a_i$ of degree $d$ is:

\begin{align}
\mathrm{LPM}^*_{i}(\boldsymbol{\varsigma}, \tau, d, S) &= \frac{1}{|A|} \sum_{s \in S} q(s) \sum_{a_j \in A} \varsigma(a_j) \notag \\
    &\quad \times \Big(\max(0, \tau - r(a_i, a_j, s))\Big)^d
\label{eq:lpm_val}
\end{align}
measuring how often the reward of action $i$ falls below the threshold $\tau$, dependent on the opponents action $a_j$.

The co-LPM (CLPM) is analogous to the co-variance between actions $(a_i, a_j)$, but instead defined in terms of LPMs. The CLPM between action $a_i$ and $a_j$ is:

\begin{align}
    \mathrm{CLPM}^*_{i, j}(\boldsymbol{\varsigma}, \tau, d, S) &= \frac{1}{|A|} \sum_{s \in S} q(s) \sum_{a_k \in A} \varsigma(a_k) \notag \\
    &\quad \times \Big(\max(0, \tau - r(a_i, a_k, s)\Big)^{d-1} \notag \\
    &\quad \times \Big(\tau - r(a_j, a_k, s)\Big)
\end{align}
which measures how the reward of action $a_i$ varies below the threshold $\tau$ in relation to the difference between the threshold and the reward of $a_j$. These values are weighted by the probability $\varsigma$ of the action being selected by the opponent. The risk matrix $\boldsymbol{\Sigma}^{\mathrm{LPM}}$ is then an $|A| \times |A|$ matrix: 
%with entries $\Sigma^{LPM}_{i,j}(\tau, p) = LPM^{*}_{i, \boldsymbol{\varsigma, \tau, p}}$ if $i = j$ and $\Sigma^{LPM}_{i,j}(\tau, p) = CLPM^*_{i, j, \boldsymbol{\varsigma}, \tau, p}$ if $i \neq j$.
\[
\boldsymbol{\Sigma}^{\mathrm{LPM}} = 
\begin{pmatrix}
\mathrm{LPM}^{*}_{1} & \mathrm{CLPM}^{*}_{1, 2} & \cdots & \mathrm{CLPM}^{*}_{1, |A|} \\
\mathrm{CLPM}^{*}_{2, 1} & \mathrm{LPM}^{*}_{2} & \cdots & \mathrm{CLPM}^{*}_{2, |A|} \\
\vdots & \vdots & \ddots & \vdots \\
\mathrm{CLPM}^{*}_{|A|, 1} & \mathrm{CLPM}^{*}_{|A|, 2} & \cdots & \mathrm{LPM}^{*}_{|A|} \\
\end{pmatrix}
\]
Finally, we can define the $\operatorname{Risk}$ of the mixed-strategy $\boldsymbol{\sigma}$ as:

\begin{align}
    \operatorname{Risk}(\boldsymbol{\sigma}, \boldsymbol{\varsigma}, S) &= \sum_{k=1}^{|A|} \sum_{n=1}^{|A|} \sigma(a_k) \sigma(a_n) \mathrm{CLPM}^*_{k, n}(\boldsymbol{\varsigma}, \tau, d, S)\\ 
    &= \boldsymbol{\sigma}^T \cdot \boldsymbol{\Sigma}^{\mathrm{LPM}} \cdot \boldsymbol{\sigma}
\end{align}

\textbf{Properties of $\boldsymbol{\Sigma}^{\mathrm{LPM}}$}

The asymmetry of $\boldsymbol{\Sigma}^{\mathrm{LPM}}$ poses a concern for mathematical optimisation, as symmetry is a requirement for quadratic programming. Various techniques to render $\boldsymbol{\Sigma}^{\mathrm{LPM}}$ symmetric have been proposed with different benefits \cite{cumova2011symmetric, estrada2004mean, nawrocki1992characteristics}. We consider three of the most widely used symmetrization approaches, to show DRAE is independent of the specific symmetrization used. All three of these approaches are suitable and the exact choice of approach will be environment dependent. The approaches are detailed below.

%While each of these symmetrization processes have their own limitations, we demonstrate empirically in \cref{secResults} that these approximations are all effective for reducing the true downside risk of strategy profiles. 

\textit{Rho \cite{nawrocki1992characteristics}} - This approach expresses CLPM in terms of the individual LPM values per action, and the correlation coefficient between the utility vectors of the corresponding actions. The symmetric CLPM values are:

\begin{align}
CLPM_{i,j}=CLPM_{j,i} = \Big(LPM_i \times LPM_j\Big)^{\frac{1}{d}} \rho_{i,j}
\end{align}
where $\rho_{i,j}$ is the correlation coefficient between the utility vectors of actions $i$ and $j$. The main downside of this approach is that it relies on the symmetric correlations, through $\rho_{i,j}$, rather than the partial correlations of the action utilities.

\textit{Dual \cite{estrada2008mean}} - Instead of focusing on the full symmetric correlations, as in \textit{rho}, the \textit{dual} approach considers only the actions that have utilities which both fall below the threshold:

\begin{align}
    CLPM_{i,j} = \frac{1}{|A|} \sum_{s \in \mathcal{S}}q(s)\sum_{a_k \in \mathcal{A}} \bigg[\Big(\operatorname{max}(0, \tau - u(a_i, a_k, s)\Big)^{d-1} \\ \times \Big(\operatorname{max}(0, \tau - u(a_j,a_k,s)\Big)^{d-1}\bigg]^{\frac{1}{d-1}}
\end{align}

The main limitation is, by only including values where both actions fall below the threshold, we lose information relating to only the secondary action moving above the threshold. 

\textit{Transpose \cite{cumova2011symmetric}} - The final approach leverages a standard algebraic manipulation. Specifically, we can decompose the risk matrix into a symmetric and anti-symmetric part as: 

\begin{align}
\Sigma^{LPM} &= \frac{1}{2}(\Sigma^{LPM} + \Sigma^{LPM,T}) + \frac{1}{2}(\Sigma^{LPM} - \Sigma^{LPM, T}) \\ &= \Sigma^{LPM}_s + \Sigma^{LPM}_a
\end{align}

If we examine the quadratic form of the purely anti-symmetric matrix noting that, due to its anti-symmetric nature, $\Sigma^{LPM, T}_a = -\Sigma^{LPM}_a$:

\begin{align}
    q = \mathbf{x}^T \Sigma^{LPM}_a \mathbf{x} &= (\mathbf{x}^T \Sigma^{LPM, T}_a \mathbf{x})^T \\
    &= -(\mathbf{x}^T \Sigma^{LPM}_a \mathbf{x})^T = -q
\end{align}
which implies that $q = 0$. Finally, we have that:
\begin{align}
    \mathbf{x}^T \Sigma^{LPM} \mathbf{x} &= \mathbf{x}^T(\Sigma^{LPM}_s + \Sigma^{LPM}_a)\mathbf{x} \\
    &= \mathbf{x}^T \Sigma^{LPM}_s \mathbf{x} + \mathbf{x}^T \Sigma^{LPM}_a \mathbf{x} = \mathbf{x}^T \Sigma^{LPM}_s \mathbf{x}
\end{align}

Therefore, as our optimisation formulation solves a quadratic programme (QP), we can achieve the equivalent solution for the original asymmetric risk matrix by optimising over a symmetric transformed matrix. This formulation however does not guarantee that $\Sigma_s^{LPM}$ is positive-definite (PD), which is required for uniqueness under a QP. If not PD, we utilise a simple transformation \cite{higham1988computing} to find the nearest PD matrix in terms of the Frobenius norm - in practice this approximation does not impact performance significantly.

\textbf{Optimisation}
Based on the above, we define our utility function $u$ as:

\begin{align}\label{eqDRAEUtility}
u(\boldsymbol{\sigma}, \boldsymbol{\varsigma}, \boldsymbol{M}, S) = \mathrm{ER}(\boldsymbol{\sigma}, \boldsymbol{\varsigma}, \boldsymbol{M}, S) - \gamma \operatorname{Risk}(\boldsymbol{\sigma}, \boldsymbol{\varsigma}, S)
\end{align}
where $\gamma$ is a parameter capturing the risk focus of the player. The resulting best-response (BR) map is:

\begin{equation} \label{eq:br-map}
\begin{aligned}
\boldsymbol{\sigma}^*(\boldsymbol{\varsigma}) \in \text{argmax}_{\boldsymbol{\sigma}}u(\boldsymbol{\sigma}, \boldsymbol{\varsigma}, \boldsymbol{M}, S) \\
\text{s.t. } \sigma(a) \geq \epsilon  \text{  }, \forall a \in A, \boldsymbol{\sigma}^{T}\textbf{1} = 1
\end{aligned}
\end{equation}
where, under conditions set out by \cite{cumova2011symmetric, estrada2004mean, nawrocki1992characteristics}, we have a quadratic programming problem due to the quadratic term $\boldsymbol{\sigma}^T \cdot \boldsymbol{\Sigma}^{\mathrm{LPM}} \cdot \boldsymbol{\sigma}$ and the linear constraints ($\epsilon >$ 0). The programme finds $\boldsymbol{\sigma}^*$ such that the utility is maximised, whilst ensuring the usual probability constraints (non negative probabilities and that the probabilities sum to one). 

\subsection{Solution concept}

Based on \cref{eq:br-map}, we can define our novel solution concept DRAE:

\begin{definition}[DRAE]
    A strategy profile $\{\boldsymbol{\sigma}, \boldsymbol{\varsigma}\}$ is a downside risk-aware equilibrium if both $\boldsymbol{\sigma}$ and $\boldsymbol{\varsigma}$ are risk-aware responses, in that they satisfy Eq. \ref{eq:br-map}, to each other.
\end{definition}

Importantly, for a given expected return,  the DRAE corresponds to the minimum LPM solution \cref{secMinLPM}, and furthermore, is guaranteed to exist \cref{appendixExistence}.

\subsubsection{Minimum LPM Solution}\label{secMinLPM}

For a given expected return, DRAE corresponds to the minimum risk solution.

The best response map that we optimise for in \cref{eq:br-map} has the same solution as the following:
\begin{equation}
\begin{aligned}
\boldsymbol{\sigma}^*(\boldsymbol{\varsigma}) \in \text{argmin}_{\boldsymbol{\sigma}}  \boldsymbol{\sigma}^{T} \cdot \boldsymbol{\Sigma}^{\text{LPM}} \cdot \boldsymbol{\sigma} \\
\text{s.t. } \boldsymbol{\sigma}^{T} \cdot \boldsymbol{M} \cdot \boldsymbol{\varsigma} \geq \mu_{\text{b}}\\
\sigma(a) \geq 0 \text{  } \forall a \in A\\
\boldsymbol{\sigma}^{T}\textbf{1} = 1\\
\boldsymbol{\varsigma}^{T}\textbf{1} = 1
\end{aligned}
\end{equation}
where $\mu_{\text{b}} \in \mathbb{R}$ is a lower bound on the expected return. In other words, given a pre-defined expected return value of $\mu_b$, $\boldsymbol{\sigma}^*(\boldsymbol{\varsigma})$ also minimises the risk, $\boldsymbol{\sigma}^T \cdot \boldsymbol{\Sigma}^{\text{LPM}} \cdot \boldsymbol{\sigma}$

\begin{proof}
\cite{merton1972analytic} shows, as $\boldsymbol{\Sigma}^{\text{LPM}}$ is enforced to be Positive Definite (PD) we have 
\begin{equation}
\begin{aligned}
\boldsymbol{\sigma}^*(\boldsymbol{\varsigma}) \in \text{argmin}_{\boldsymbol{\sigma}}  \boldsymbol{\sigma}^{T} \cdot \boldsymbol{\Sigma}^{\text{LPM}} \cdot \boldsymbol{\sigma} \\
\text{s.t. } \boldsymbol{\sigma}^{T} \cdot \boldsymbol{M} \cdot \boldsymbol{\varsigma} \geq \mu_{\text{b}}\\
\sigma(a) \geq \epsilon \text{  } \forall a \in A\\
\boldsymbol{\sigma}^{T}\textbf{1} = 1
\end{aligned}
\end{equation}

\noindent which can be rewritten through a Lagrange multiplier as
\begin{equation}
\begin{aligned}
\boldsymbol{\sigma}^*(\boldsymbol{\varsigma}) \in \text{argmin}_{\boldsymbol{\sigma}}  \boldsymbol{\sigma}^{T} \cdot \Sigma^{\text{LPM}} \cdot \boldsymbol{\sigma} - \gamma\Big(\boldsymbol{\sigma}^{T} \cdot \boldsymbol{M} \cdot \boldsymbol{\varsigma}\Big)\\
\text{s.t. } 
\sigma(a) \geq \epsilon \text{  } \forall a \in A\\
\boldsymbol{\sigma}^{T}\textbf{1} = 1
\end{aligned}
\end{equation}

\noindent which can be equivalently expressed as,
\begin{equation}
\begin{aligned}
\boldsymbol{\sigma}^* \in \text{argmin}_{\boldsymbol{\sigma}}  -\Big(\boldsymbol{\sigma}^{T} \cdot \boldsymbol{M} \cdot \boldsymbol{\varsigma} - \lambda \boldsymbol{\sigma}^{T} \cdot \Sigma^{\text{LPM}} \cdot \boldsymbol{\sigma}\Big)\\
\text{s.t. } 
\sigma(a) \geq \epsilon \text{  } \forall a \in A\\
\boldsymbol{\sigma}^{T}\textbf{1} = 1
\end{aligned}
\end{equation}

where $\lambda = \frac{1}{\gamma}$, which can be rewritten as \cref{eq:br-map}.

\end{proof}

\subsubsection{DRAE Existence}\label{appendixExistence}

To demonstrate the existence of DRAE, we show that DRAE satisfies the four conditions required by Kakutani's fixed-point theorem, 
following the proof from \cite{slumbers2023game}.

For any finite N-player game where each player $i$ has a finite $k$ number of pure strategies, $A^i = \{a^i_1, ..., a^i_k\}$, a DRAE exists.

\begin{proof}
%We base our proof on Kakutani's Fixed Point Theorem, and show that DRAE satisfies the four conditions.

We define our best-response function as $B_i(\boldsymbol{\sigma}_{-i}) = \text{argmax}_{\boldsymbol{\sigma}} u^i(\boldsymbol{\sigma}, \boldsymbol{\sigma}_{-i})$ where $u_i$ is defined as in Eq. (16) and by definition $\boldsymbol{\sigma}$ must satisfy all of the properties of a proper mixed-strategy, and the best-response correspondence is $B: \Delta \rightarrow \Delta^N$ such that for all $\boldsymbol{\sigma} \in \Delta$, we have:

\begin{equation}
    B(\boldsymbol{\sigma}) = [B_i(\boldsymbol{\sigma}_{-i})]_{i\in N}
\end{equation}

We show that $B(\boldsymbol{\sigma})$ satisfies the conditions of Kakutani's Fixed Point Theorem

\begin{enumerate}
    \item \textit{$\Delta$ is compact, convex and non-empty.}
    
    By definition 
    \begin{equation}
        \Delta = \Pi_{i \in N} \Delta_i
    \end{equation}
    where each $\Delta_i = \{a | \sum_j a_j = 1 \}$ is a simplex of dimension $|A^i| - 1$, thus each $\Delta_i$ is closed and bounded, and thus compact. Their product set is also compact. 
    
    \item \textit{$B(\boldsymbol{\sigma})$ is non-empty.} 
    
    By the definition of $B_i(\boldsymbol{\sigma}_{-i})$ where $\Delta_i$ is non-empty and compact, and $u^i$ is a quadratic (as demonstrated in Sec. 3.2) and hence a polynomial function in $a$. As all polynomial functions are continuous, we can invoke Weirstrass's Extreme Value Theorem which states:
    
    If a real valued-function $f$ is continuous on the closed interval $[a,b]$, then $f$ must attain a maximum and a minimum, each at least once. That is, there exist numbers $c$ and $d$ in $[a,b]$ such that:
    \begin{equation*}
        f(c) \geq f(x) \geq f(d) \quad \forall x \in [a,b]
    \end{equation*}
    
    Therefore, as $\Delta_i$ is non-empty and compact and $u^i$ is continuous in $a$, $B_i(\boldsymbol{\sigma}_{-i})$ is non-empty, and therefore $B(\boldsymbol{\sigma})$ is also non-empty.
    
    \item \textit{$B(\boldsymbol{\sigma})$ is a convex-valued correspondence}.
    
    Equivalently, $B(\boldsymbol{\sigma}) \subset \Delta$ is convex if and only if $B_i(\boldsymbol{\sigma}_{-i})$ is convex for all $i$. 
    
    In order to show that $B_i(\boldsymbol{\sigma}_{-i}$) is convex for all $i$, we instead show that the Quadratic Programme defined by Eq. (17) is a special case of convex optimisation under certain conditions, and therefore by definition has a feasible set which is a convex set. 
    
    A \textit{convex optimisation problem} is one of the form,
    
    \begin{equation} 
    \begin{aligned}
    \text{minimize} \quad &f_0(x)\\
    \text{s.t. } &f_i(x) < 0, \text{  } i=1,...,m\\
    &a_i^{T}x = b_i, \quad i=1,...,p
    \end{aligned}
    \end{equation}
    
    where $f_0, ..., f_m$ are convex functions. The requirements for a problem to be a convex optimisation problem are:
    
    \begin{enumerate}
        \item the objective function must be convex
        \item the inequality constraint functions must be convex
        \item the equality constraint functions $h_i(x)=a_i^{T}x = b_i$ must be affine
    \end{enumerate}
    
    We note that a quadratic form $\mathbf{x}^{T} \boldsymbol{A} \mathbf{x}$ is convex if $\boldsymbol{A}$ is positive semi-definite, and strictly convex if $\boldsymbol{A}$ is positive definite. In our constrained optimisation, the quadratic term $\boldsymbol{\sigma}^{T} \boldsymbol{\Sigma}^{\text{LPM}} \boldsymbol{\sigma}$ is always guaranteed to be at least convex as $\boldsymbol{\Sigma}^{\text{LPM}}$, the LPM matrix, is always at least PSD as we enforce it to be this way (see the final paragraph of page 3, below Eq. (15)). Therefore, our objective function is convex. Additionally, our inequality constraint functions are convex (because they are linear) and our equality constraint function is affine. Therefore, our Quadratic Programme is an instance of a convex optimisation problem.
    Importantly, the feasible set of a convex optimisation problem is convex, since it is the intersection of the domain of the problem: $\mathcal{D} = \bigcap_{i=0}^m \textbf{dom} f_i$ which itself is a convex set. 
    
    Therefore, for all members of the feasible set $x, y \in B_i(\boldsymbol{\sigma}_{-i})$ and all $\theta\in [0,1]$ we have that $\theta x + (1-\theta)y \in S$ and we have a convex-valued correspondence.
    
    \item \textit{$B(\boldsymbol{\sigma})$ has a closed graph.}
    
    Suppose to obtain a contradiction, that $B(\boldsymbol{\sigma})$ does not have a closed graph. Then, there exists a sequence ($\boldsymbol{\sigma}^n, \hat{\boldsymbol{\sigma}}^n) \rightarrow (\boldsymbol{\sigma}, \hat{\boldsymbol{\sigma}})$ with $\hat{\boldsymbol{\sigma}}^n \in B(\boldsymbol{\sigma}^n)$, but $\hat{\boldsymbol{\sigma}} \notin B(\boldsymbol{\sigma})$, i.e. there exists some $i$ such that $\hat{\boldsymbol{\sigma}}_i \notin B_i(\boldsymbol{\sigma}_{-i})$. This implies that there exists some $\boldsymbol{\sigma}_i^{\prime} \in \Delta_i$ and some $\epsilon > 0$ such that 
    
    \begin{equation}
        u_i(\boldsymbol{\sigma}_i^{\prime}, \boldsymbol{\sigma}_{-i}) > u_i(\hat{\boldsymbol{\sigma}}_i, \boldsymbol{\sigma}_{-i}) + 3\epsilon.
    \end{equation}
    
    By the continuity of $u_i$ and the fact that $\boldsymbol{\sigma}_{-i}^n \rightarrow \boldsymbol{\sigma}_{-i}$, we have for sufficiently large $n$,
    
    \begin{equation}
        u_i(\boldsymbol{\sigma}_i^{\prime}, \boldsymbol{\sigma}_{-i}^n) \geq u_i(\boldsymbol{\sigma}_i^{\prime}, \boldsymbol{\sigma}_{-i}) - \epsilon.
    \end{equation}
    
    and combining the preceding two relations we obtain
    
    \begin{equation}
        u_i(\boldsymbol{\sigma}_i^{\prime}, \boldsymbol{\sigma}_{-i}^n) > u_i(\hat{\boldsymbol{\sigma}}_i, \boldsymbol{\sigma}_{-i}) + 2\epsilon \geq u_i(\hat{\boldsymbol{\sigma}}_i^n, \boldsymbol{\sigma}_{-i}^n) + \epsilon
    \end{equation}
    
    where the second relation follows from the continuity of $u_i$. This contradicts the assumption that $\hat{\boldsymbol{\sigma}}_i^n \in B(\boldsymbol{\sigma}_{-i}^n)$ and completes the proof.
\end{enumerate}

Therefore, $B(\boldsymbol{\sigma})$ satisfies the conditions of Kakutani's Fixed Point Theorem, and therefore if $\boldsymbol{\sigma}^* \in B(\boldsymbol{\sigma}^*)$ then $\boldsymbol{\sigma}^*$ is an equilibrium.
\end{proof}

\subsection{Solver}
Stochastic Fictitious Play (SFP) \cite{fudenberg1993learning} is used as our primary solver for DRAE over $\boldsymbol{M}$. SFP is a learning process where players choose a BR to others time-average strategies. In SFP, $n \geq 2$ players repeatedly play a $n-$player NFG. The state variable $Z_t \in \Delta_A$, with $Z_t^i = \frac{1}{t}\sum_{u=t}^t \boldsymbol{\sigma}_t^i$ describes the time averages of each player's behaviour, and $\sigma_t^i \in \Delta_{A^i}$ represents the observed strategy of player $i$ at time-step $t$. Each player best responds to the time-average strategy of their opponent, $Z_t^{-i}$, by maximising a perturbed utility function $\bar{u}$:

\begin{equation}\label{eqBR}
\begin{split}
    \boldsymbol{\sigma}_{t+1}^i &= \text{argmax}_{\boldsymbol{\sigma}} \bar{u}
    = \text{argmax}_{\boldsymbol{\sigma}} u^i(\boldsymbol{\sigma}, Z_t^{-i}, \boldsymbol{M}) - \lambda v^i(\boldsymbol{\sigma})
\end{split}
\end{equation}
where $v^i(\boldsymbol{\sigma}) : \Delta_A \rightarrow \mathbb{R}$ perturbs $u^i$ such that it is strictly concave (unique global maximum) whilst applying greater than zero probability to all actions. We provide a proof of convergence of SFP in \cref{appendixSFP}.

%Note that for SFP we require a stronger notion of convergence in observed strategies $\boldsymbol{\sigma}_t^i$ rather than in the beliefs $Z_t^i$ (which is all that vanilla FP requires), but by definition a converged final $\boldsymbol{\sigma}_t^i$ is a DRAE. 

\subsection{Connection to existing equilibrium concepts}

It is trivial to demonstrate that in $\cref{eqDRAEUtility}$ as $\gamma \to 0$ then the resulting BR map of $\cref{eq:br-map}$ is equivalent to the Nash solution that focuses on expected reward alone. Under certain conditions of the reward structure and/or threshold value, RAE can be recovered as a special instance of DRAE. 

\textbf{Case 1}: 
When the variance of the rewards for every action is 0, then RAE and DRAE both attain the Nash solution as both the co-variance matrix $\Sigma$ and the co-LPM matrix $\Sigma^{\text{LPM}}$ have all 0 entries - there is no risk under either approach. The threshold value $\tau$ does not matter for this case as there are no values that deviate from the mean.

\textbf{Case 2}: Consider a matrix where the entries are $\{-1, 1\}$, the table is symmetric $r(a_i, a_j) = r(a_j, a_i)$ and the variance is constant across actions (implies the same number of wins and losses for an action). We assume that $\tau$ is the mean of the payoff matrix, so $\tau=0$. First, consider the diagonal elements of $\Sigma^{LPM}$ (for DRAE) and the co-variance matrix $\Sigma$ (for RAE). The diagonal elements of $\Sigma^{LPM}$ are the LPM values for each action $i$, calculated by \cref{eq:lpm_val}. As the same amount of rewards $r(a_i, a_j)$ for $a_i$ fall below $\tau$ for every action, the LPM values are constant across all actions. By definition, the diagonal elements of $\Sigma$ are the variances of each action which are the same across all actions. For the off-diagonal elements, first note that, due to symmetry, the payoff vectors for action $i$, $\boldsymbol{M}_i$, and action $j$, $\boldsymbol{M}_j$, are either the same, $\boldsymbol{M}_i = \boldsymbol{M}_j$, or the opposite, $\boldsymbol{M}_i = -\boldsymbol{M}_j$. Now consider $CLPM_{i,j}$, if we have that $\boldsymbol{M}_i = \boldsymbol{M}_j$ then all of the terms in the sum are positive (as when $r(a_i,a_k) < \tau$ then the product is of two negative values), and we have the opposite when $\boldsymbol{M}_i = -\boldsymbol{M}_j$. However, the absolute value of the elements are always the same, therefore $CLPM_{ij} = CLPM_{ji}$ if $\boldsymbol{M}_i = \boldsymbol{M}_j$ and $CLPM_{ij} = -CLPM_{ji}$ if $\boldsymbol{M}_i = -\boldsymbol{M}_j$. Furthermore, $|CLPM_{ij}| = |LPM_{i}| \; \forall i,j$ as the same number of terms are raised to the power of degree $d$ for both. The same argument can be made for the off-diagonal elements of the co-variance matrix $\Sigma$, however all of the terms (double the amount) contribute instead of only those that fall below $\tau$, but the signs remain consistent between $\Sigma^{LPM}$ and $\Sigma$. Due to the sign consistency, and that the diagonal and off-diagonal elements are the same (within their own matrices, not across the matrices), the relative risk valuations across DRAE and RAE are the same, just scaled. Therefore, for suitable $\gamma_{RAE}$ and $\gamma_{DRAE}$ the solutions are equivalent.

%First consider the LPM value for each action and the variance for each action - by definition they are constant for each action. Second, consider the co-variance values, they are the same magnitude as the variance value but can have altering sign. The co-LPM matrix has the same property that the magnitude of all of the entries is the same as the LPM, but the sign alters. Notably, the signs are the same across the co-variance and the co-LPM matrix. Therefore, the relative risk values for RAE and DRAE are the same across the actions, leading to the same final strategy.

\begin{figure}
    \centering
    \includegraphics[width=.85\linewidth]{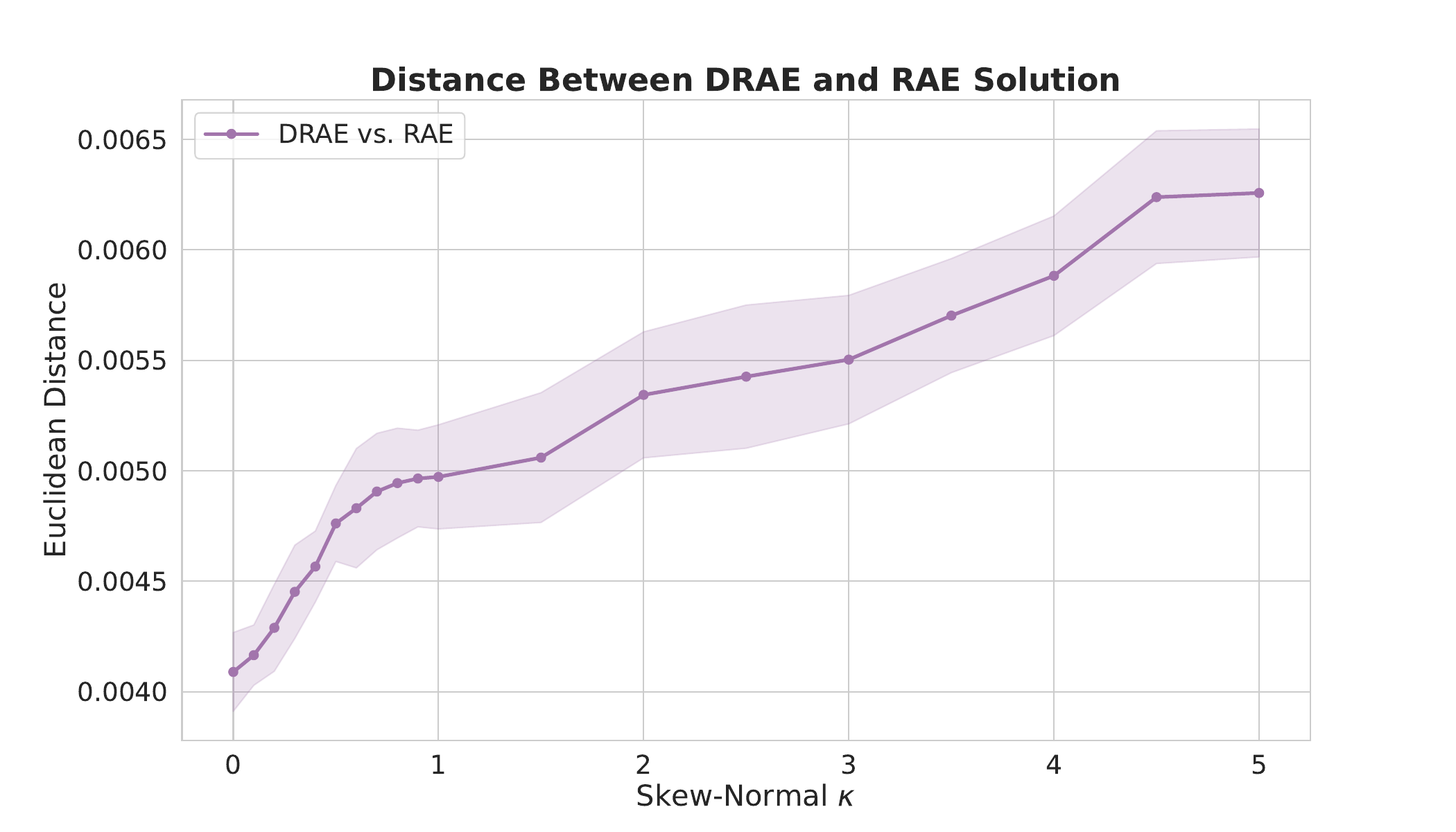}
    \caption{The normalised Euclidean distance between the DRAE and RAE equilibrium mixed strategies (over 50 seeds) as the skewness of the payoff distribution increases.}
    \label{fig:drae_kappa}
\end{figure}

However, in practice we are unlikely to come across payoff matrices that meet the above criteria. Instead, we provide an illustrative example that demonstrates how the distance between the RAE and DRAE solutions increase as we introduce more skewness $\kappa$ into payoff matrices. We generate random payoff matrices sampled from a standard skew-normal distribution for a game with 100 actions, and visualise the distances between the RAE and DRAE equilibrium solutions (\cref{fig:drae_kappa}). \footnote{At $\kappa=0$ the distance is not exactly zero because small differences in non-zero probabilities compound over 100 actions.} As the skew $\kappa$ increases, the distance between the solutions increases monotonically as DRAE accounts for this payoff skewness, which is ignored by RAE. In the experiments section, we show the importance of this consideration.% up to a threshold where it no longer increases at very large negative skewness.

%{
%\color{red}TODO: This seciton.

%It is trivial to see that in $\cref{eqDRAEUtility}$ as $\gamma \to 0$, we recover the Nash solution that focuses on expected reward alone. Under certain configurations of DRAE we can also demonstrate that we recover RAE as a special instance, showing DRAE can be seen as a generalisation of these existing solution concepts.
%The benefit of DRAE is being able to move beyond these limiting cases to show higher degrees of risk preferences.

%}

\subsection{Theoretical Properties}
Whilst under the large majority of reward distributions the RAE and DRAE solutions are distinct, DRAE does inherit all of the same theoretical properties of the RAE solution, as long as $\Sigma^{\text{LPM}}$ has the properties of symmetry and PD.

\begin{enumerate}
    \item For any finite $N$-player game where each player $i$ has a finite number $k$ of pure strategies, $A^i = \{a_1^i, ..., a_k^i\}$ a DRAE exists.
    \item The best-response \cref{eq:br-map} satisfies the necessary conditions for an SFP\cite{fudenberg1993learning} process, which constitutes our equilibrium solver. Specifically, the LPM risk measure is a suitable perturbation function $v^i(\boldsymbol{\sigma})$.
    \item For all possible values of the expected reward, the DRAE best-response and solution will provide the minimum possible downside risk in terms of LPM value.
\end{enumerate}

\section{Experiments}\label{secResults}

We consider three risk-focused environments: a synthetic game \cite{slumbers2023game}, an asset game \cite{alos2005asset}, and a portfolio game \cite{sadeghi2011game}. The synthetic environment is used for analysing risk in a tractable and understandable manner. We then analyse two more complex canonical games, motivated by financial and economic settings. In each case, we compare DRAE with Nash Equilibrium (NE) and RAE. %The full source code for running the experiments is available in the supplementary material. %While we focus particularly on financial applications here, DRAE has general applicability to any risk-focused environments.

\begin{figure*}[!htb]
     \centering
     \begin{subfigure}[b]{0.31\textwidth}
         \centering
         \includegraphics[width=\textwidth]{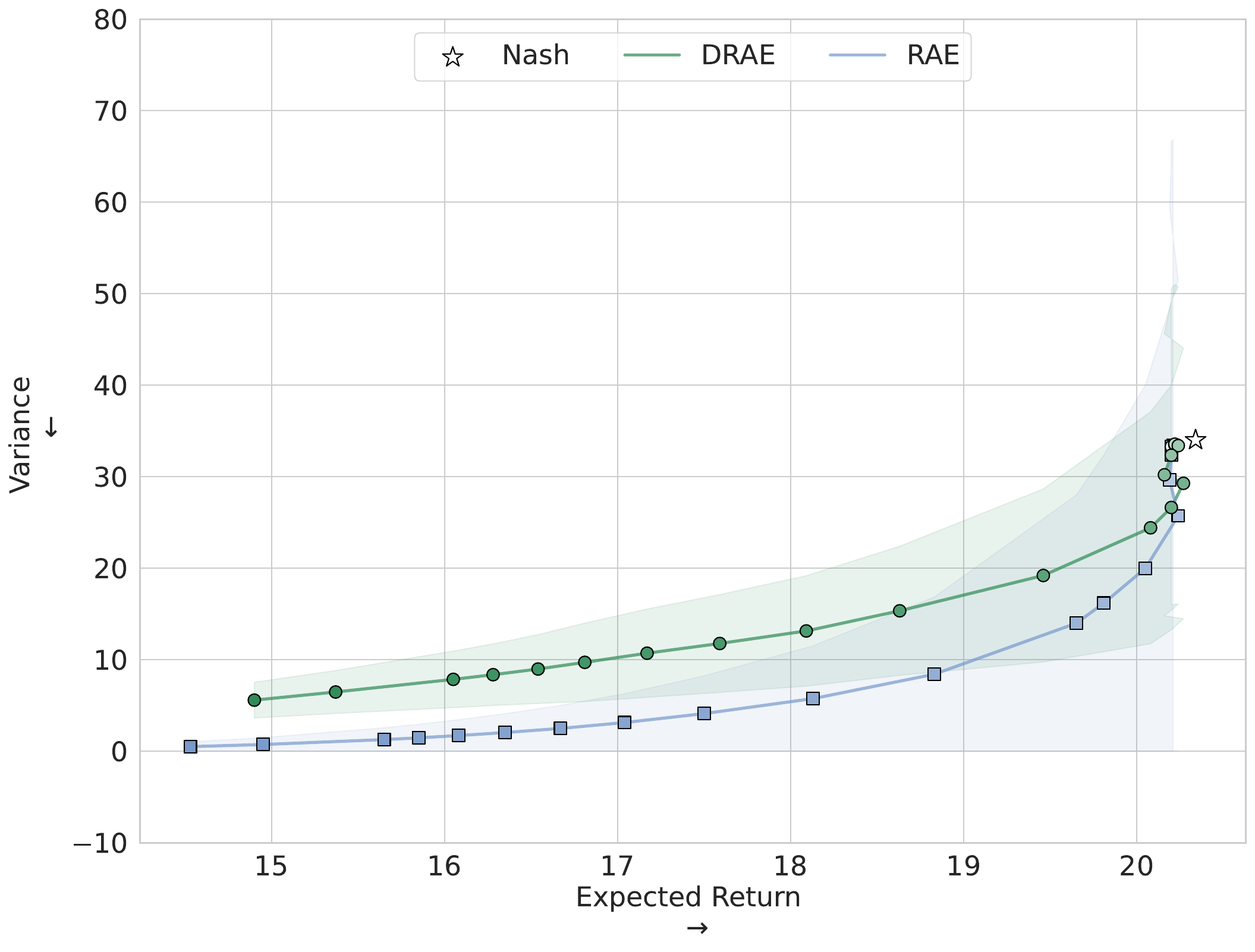}
         \caption{Synthetic: Variance}
         \label{figSyntheticVariance}
     \end{subfigure}
     \hfill
     \begin{subfigure}[b]{0.31\textwidth}
         \centering
         \includegraphics[width=\textwidth]{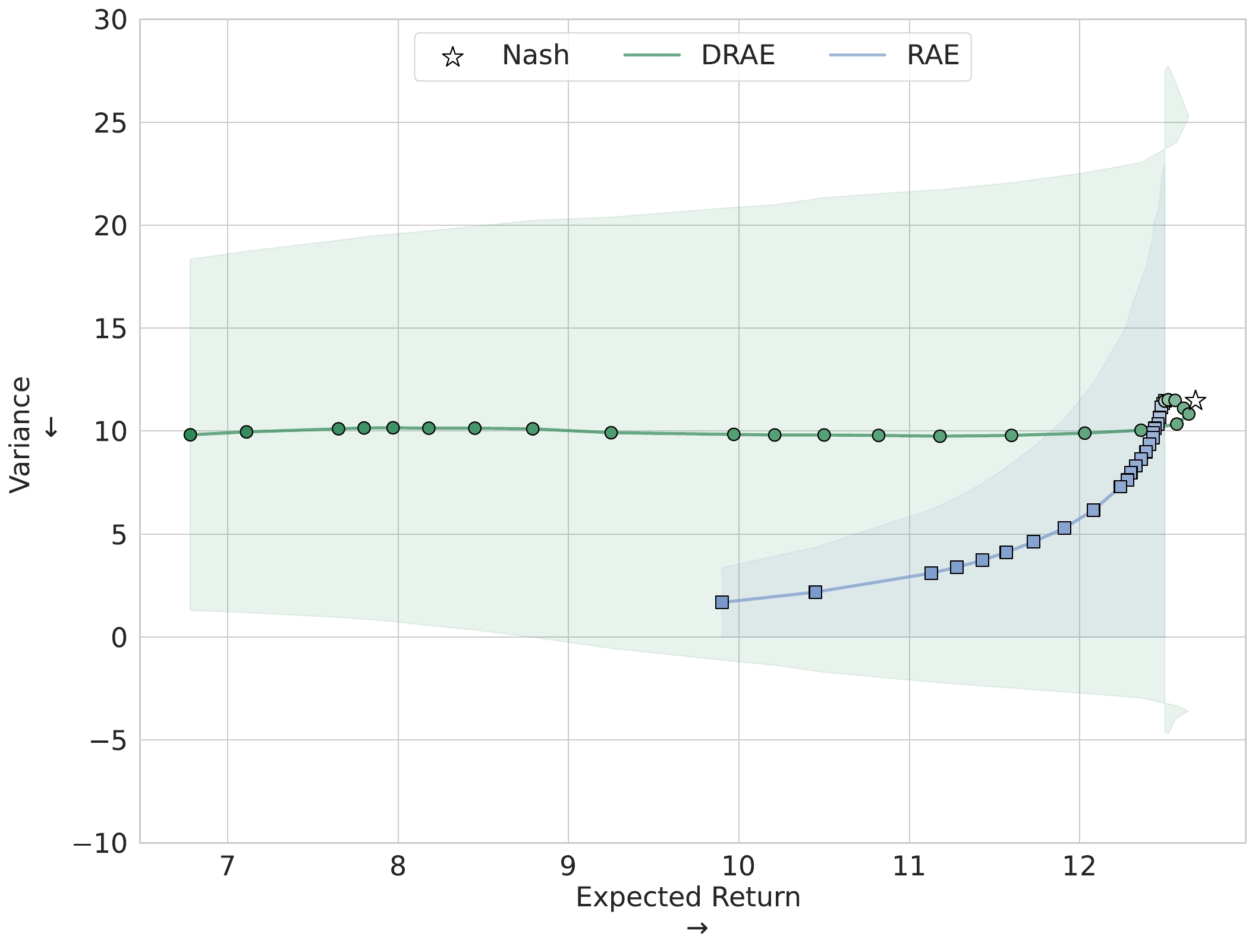}
         \caption{Asset: Variance}
         \label{figAssetVariance}
     \end{subfigure}
     \hfill
     \begin{subfigure}[b]{0.31\textwidth}
         \centering
         \includegraphics[width=\textwidth]{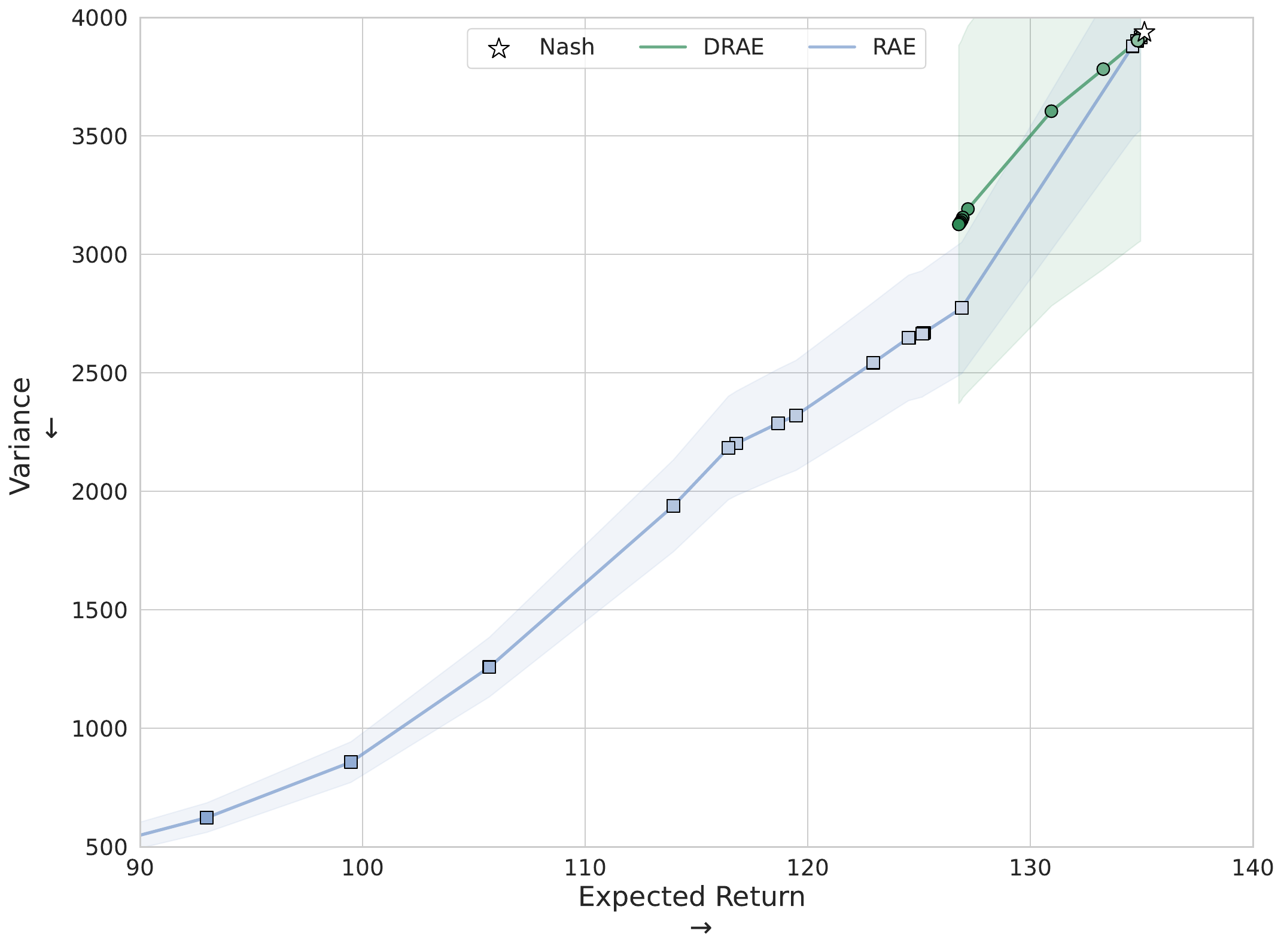}
         \caption{Portfolio: Variance}
         \label{figPPMVar}
     \end{subfigure}
     \hfill 
     \begin{subfigure}[b]{0.31\textwidth}
         \centering
         \includegraphics[width=\textwidth]{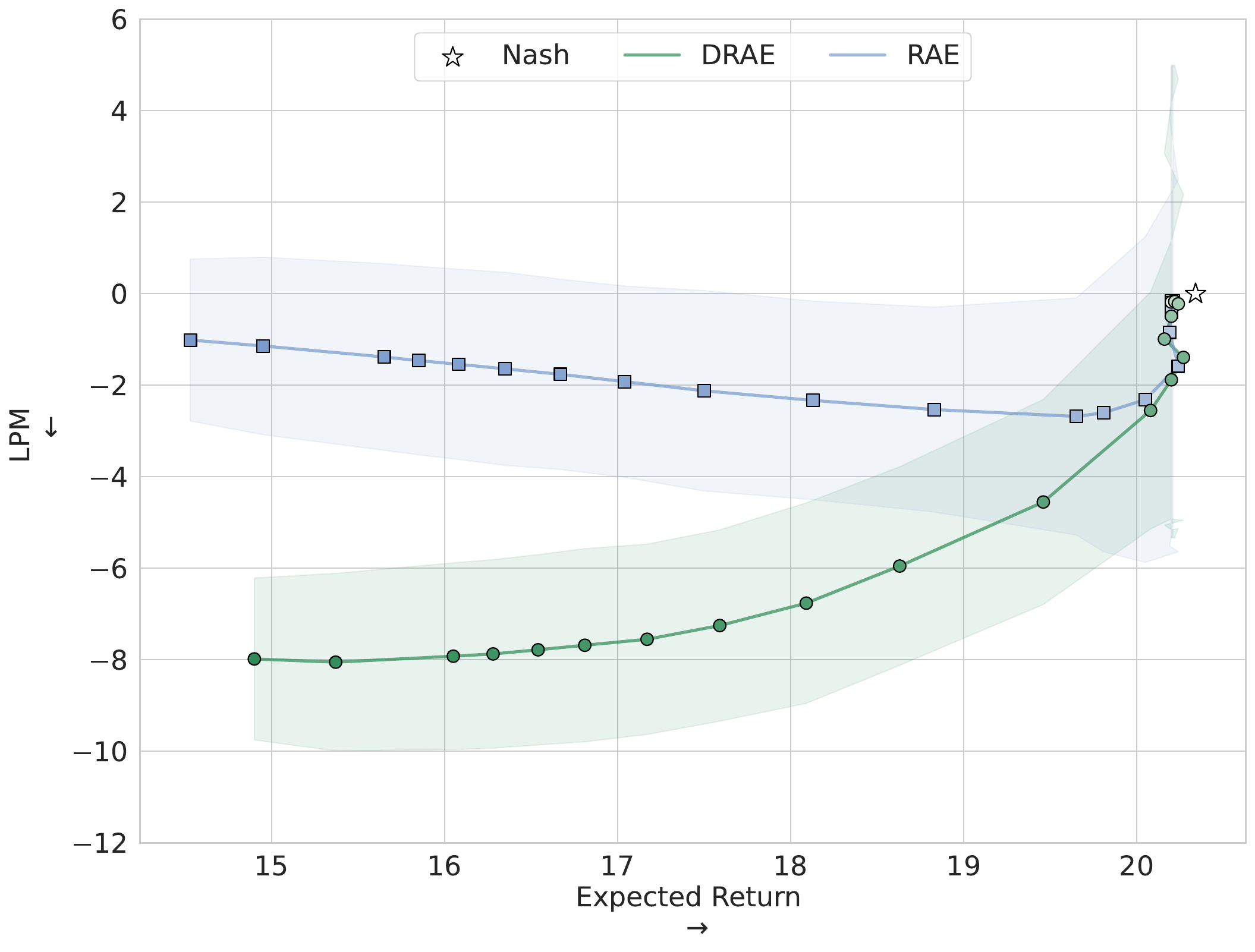}
         \caption{Synthetic: Downside risk}
         \label{figSyntheticDownside}
     \end{subfigure}
     \hfill
     \begin{subfigure}[b]{0.31\textwidth}
         \centering
         \includegraphics[width=\textwidth]{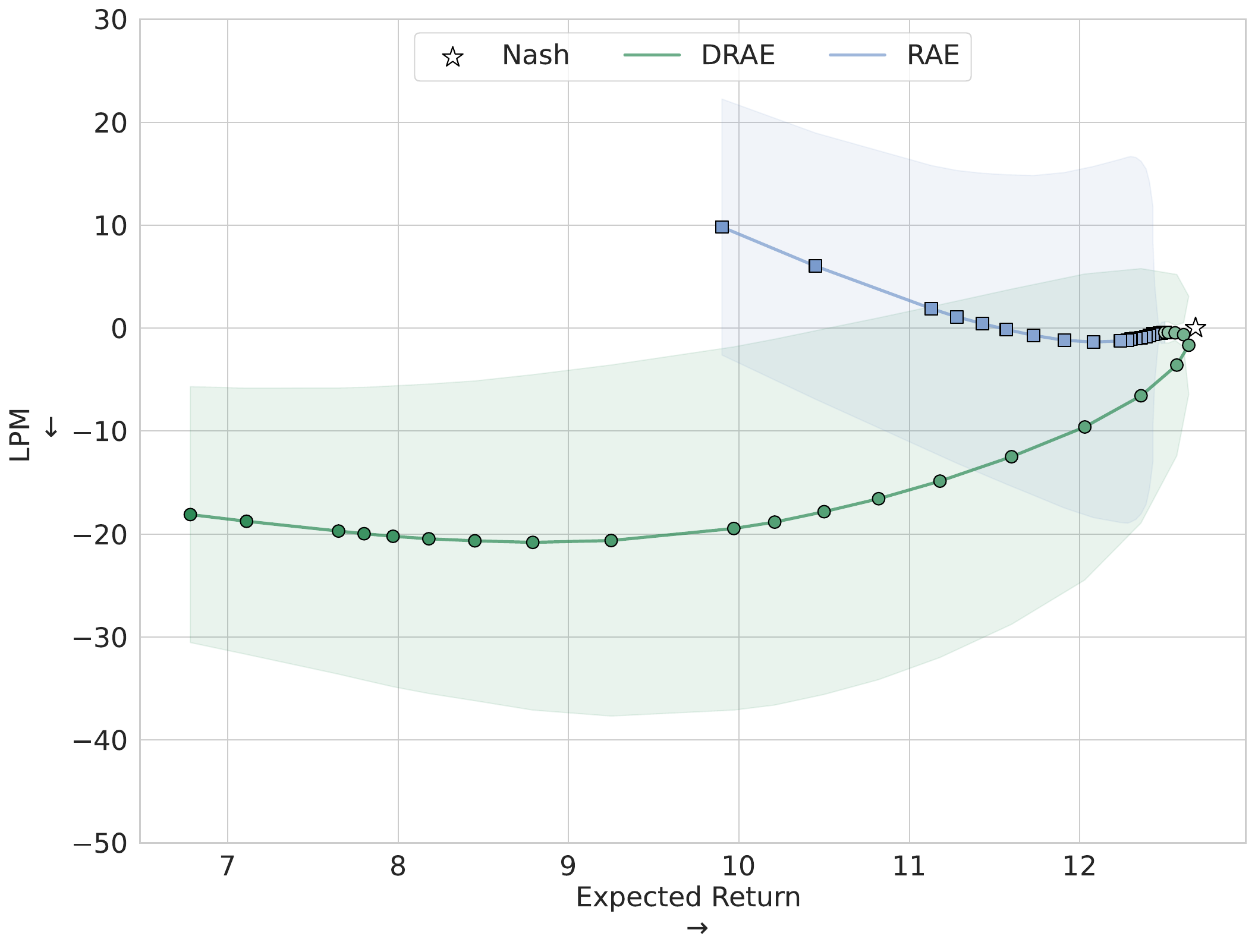}
         \caption{Asset: Downside risk}
         \label{figAssetDownside}
     \end{subfigure}
     \hfill
     \begin{subfigure}[b]{0.31\textwidth}
         \centering
         \includegraphics[width=\textwidth]{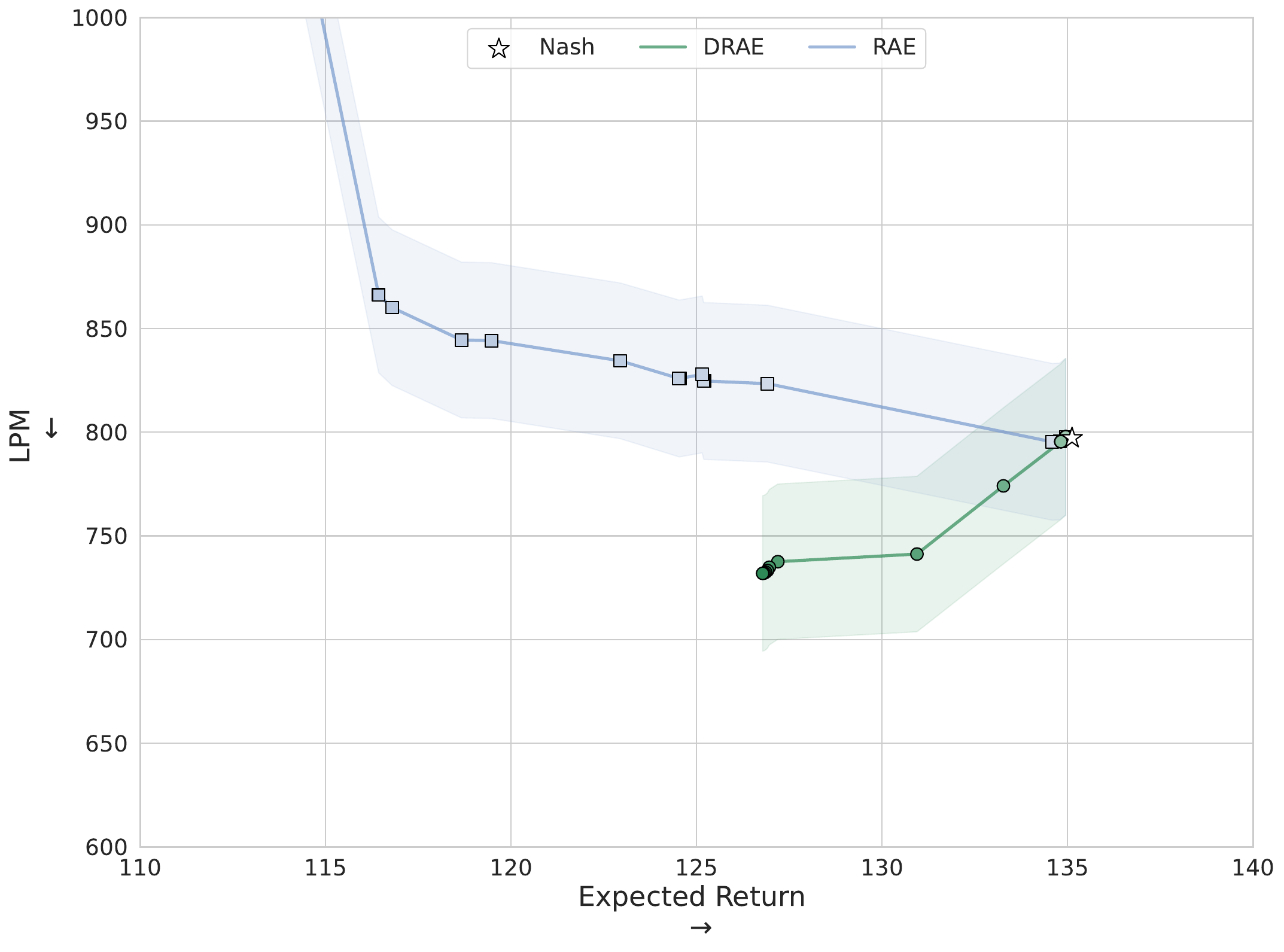}
         \caption{Portfolio: Downside risk}
         \label{figPPMDown}
     \end{subfigure}
        \caption{Variance (top row) versus Downside risk (bottom row) across the various environments (columns) for a range of gamma values $\circ$ ($\gamma$ increasing as we move right-to-left). In each case, the $x$-axis shows the expected return, while the $y$-axis is the risk measure.}
        \label{figResults}
\end{figure*}

A summary of our main results (\cref{figResults}) is provided in \cref{tblResults} demonstrating that DRAE is able to successfully minimise downside risk while achieving equivalent expected reward (to RAE) in all three environments. %This comparison is achieved by comparing the areas under the downside risk curves of RAE and DRAE over the range of expected rewards that both methods attain. 

\begin{table}[!htb]
\caption{Summary results. Normalised area under the downside risk curve (w.r.t. RAE)  across the (common) expected reward range \protect\footnotemark. A lower value indicates a lower downside risk for the same range of expected returns, therefore, the lower the better.}\label{tblResults}
\centering
\small
\begin{tabular}{@{}llll@{}}
\toprule
              & \textbf{Synthetic} & \textbf{Asset Game} & \textbf{Portfolio} \\ \midrule
\textbf{DRAE} & 0.2                & 0.23                & 0.93               \\
RAE & 1                & 1                & 1               \\
\bottomrule
\end{tabular}
\end{table}
\footnotetext{For example, in \cref{figAssetDownside} RAE and DRAE have common expected return between $[\sim 10, \sim 12.5]$, and this is the area we compare.}

\subsection{Synthetic Game}
\label{sec:synthetic}

\paragraph{Environment}
We begin our analysis of DRAE in a synthetic environment (based on \cite{slumbers2023game}), specifically designed to demonstrate differences in risk-based approaches. In this synthetic game, there are $|A_i|=100$ actions per player (here $A_i = A \forall i$), with action rewards sampled from skew-normal distributions $r \sim \mathcal{N}(\mu, \sigma^2, \kappa)$ \cite{azzalini1996multivariate}, where $\mu=0$ is the mean, $\sigma^2$ the variance, and $\kappa$ the skew. Rewards are sampled such that certain actions have rewards with high variance, low variance, high skewness, or low skewness through varying $\sigma$ and $\kappa$. 
% A skew-normal distribution is parameterised by three parameters, a mean $\mu$ and variance $\sigma^2$ (normal distribution), and a skewness parameter $\kappa$. 
Here, of the 100 actions, we randomly assign 20 action rewards to be high variance ($\sigma^2 \sim \mathcal{U}(1.5, 3.0)$) and 20 actions to have high skewness ($\kappa \sim \{-8,-7,-6, 6, 7, 8\}$). The remaining actions rewards have low variance and low skewness ($\sigma^2 \sim \mathcal{U}(0.5, 1.0)$, $\kappa \sim \mathcal{U}(-3,3))$). We note that our results are consistent when altering these parameters as long as similar relations between the high and low parameters are generally maintained. 

\paragraph{Results}
There is a unique NE solution, as identified in \cref{figSyntheticVariance}. The NE achieves a high expected reward, however, this comes with an unacceptable level of risk in terms of both variance and LPM, as reflected in \cref{figSyntheticVariance} and \cref{figSyntheticDownside}. In both of these figures, we demonstrate the performance of DRAE vs. RAE and NE over increasing values of the risk-preference parameter $\gamma$. \cref{figSyntheticVariance} demonstrates how increasing $\gamma$ leads to an expected fall in both the expected return and the variance for both RAE and DRAE, with both notably tending towards 0 variance with increasing risk-aversion. Under a more typical view of risk, RAE would appear to be the dominant strategy, as at no point does DRAE achieve a lower variance solution. However, crucially, this result is misleading when considering where the risk is coming from. In \cref{figSyntheticDownside} we show the level of downside risk for the same solutions as in \cref{figSyntheticVariance}. DRAE outperforms RAE by a large margin in terms of reducing the downside risk of their resulting equilibria. Problematically, RAE sees an increase in downside risk as it optimises for lower variance, showing the resulting strategies actually become \textit{riskier}, motivating the proposed approach.
%While RAE appears to address this risk (\cref{figSyntheticVariance}), this comes at the expense of minimising upside payoffs, as demonstrated by \cref{figSyntheticDownside}. In contrast, the proposed approach is able to balance expected reward with the true downside risk, minimising the potential for negative payoffs (\cref{figSyntheticDownside}). This is further confirmed in \cref{tblResults}.

%Through varying the risk preference $\gamma$, we demonstrate how we can alter this balance between expected reward and downside risk. The specific value of $\gamma$ to use depends on the decision-makers being modelled, with $\gamma \to \inf$ becoming more risk-averse.

\begin{figure*}[!htb]
    \centering
    \begin{minipage}{0.31\textwidth}
    \centering
    \includegraphics[width=\textwidth]{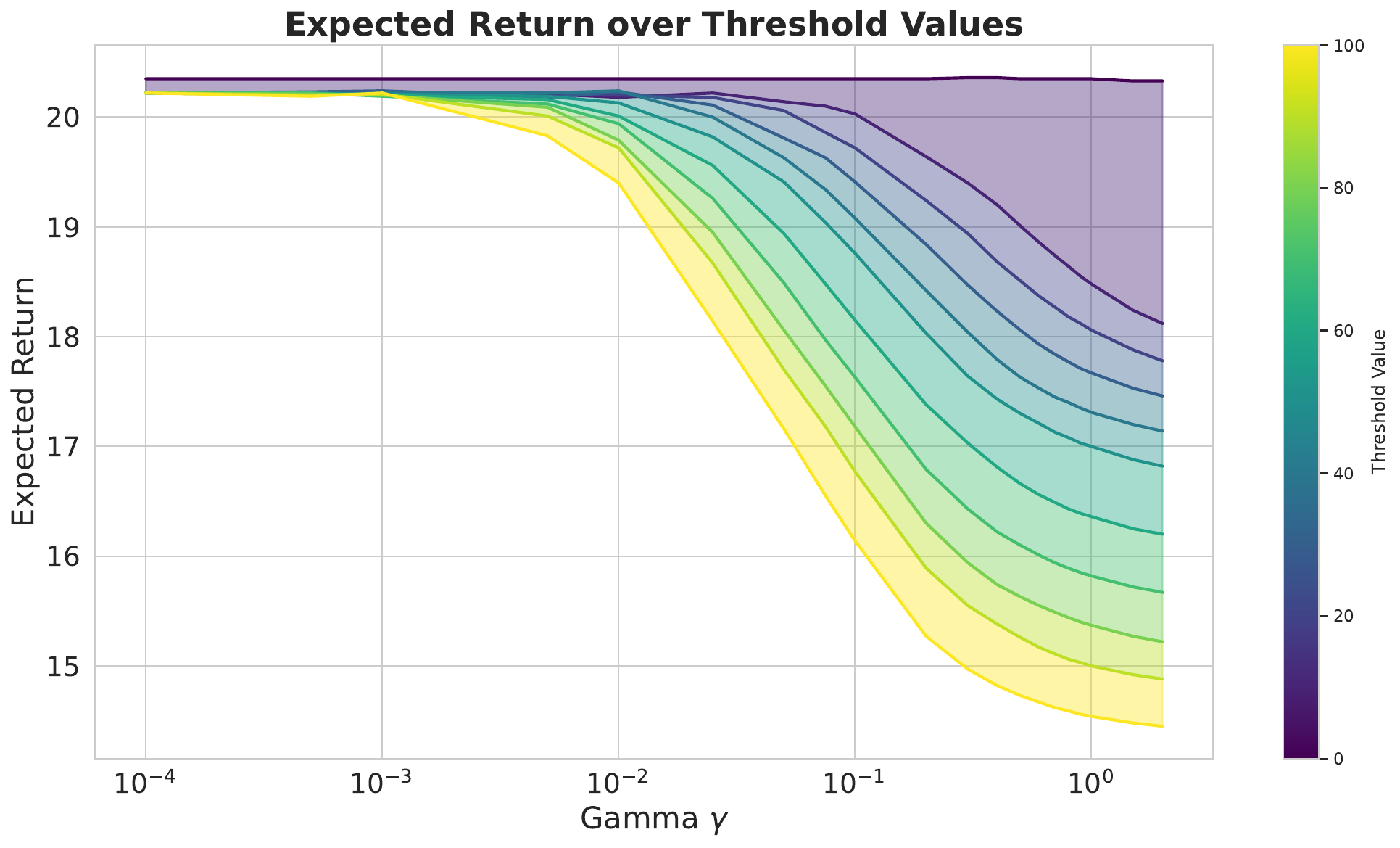}
    \caption{Comparison of the expected return value for DRAE over a large range of threshold values $\tau$ on synthetic games.}
    \label{figLowTau}
    \end{minipage}\hfill
\begin{minipage}{0.31\textwidth}
        \centering
    \includegraphics[width=\textwidth]{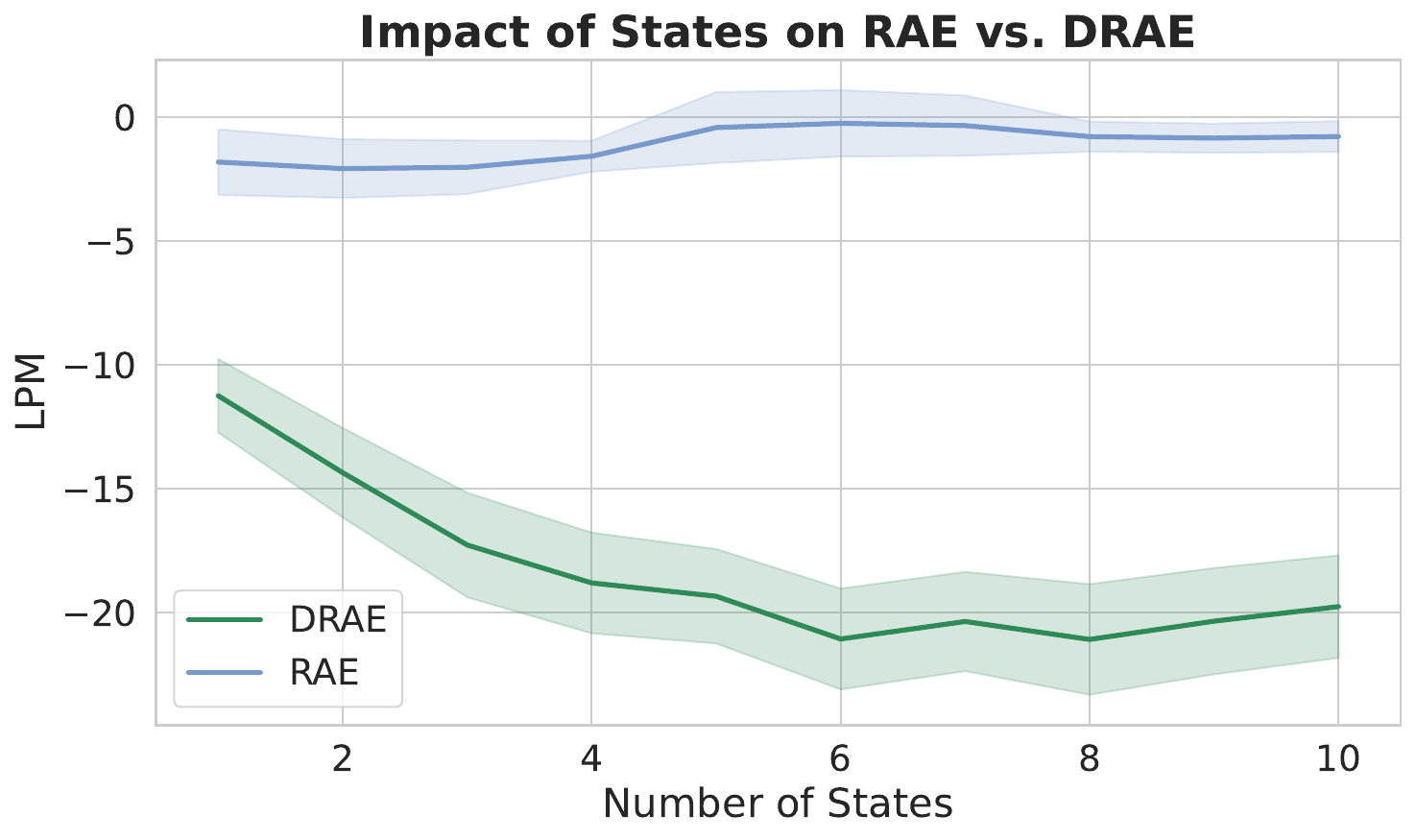}
    \caption{Equilibrium LPM under varying levels of exogenous risk (through adding more risky states to $\mathcal{S}$).}
    \label{figEndogenousRisk}
    \end{minipage}\hfill
    \begin{minipage}{0.31\textwidth}
    \centering
    \includegraphics[width=\textwidth]{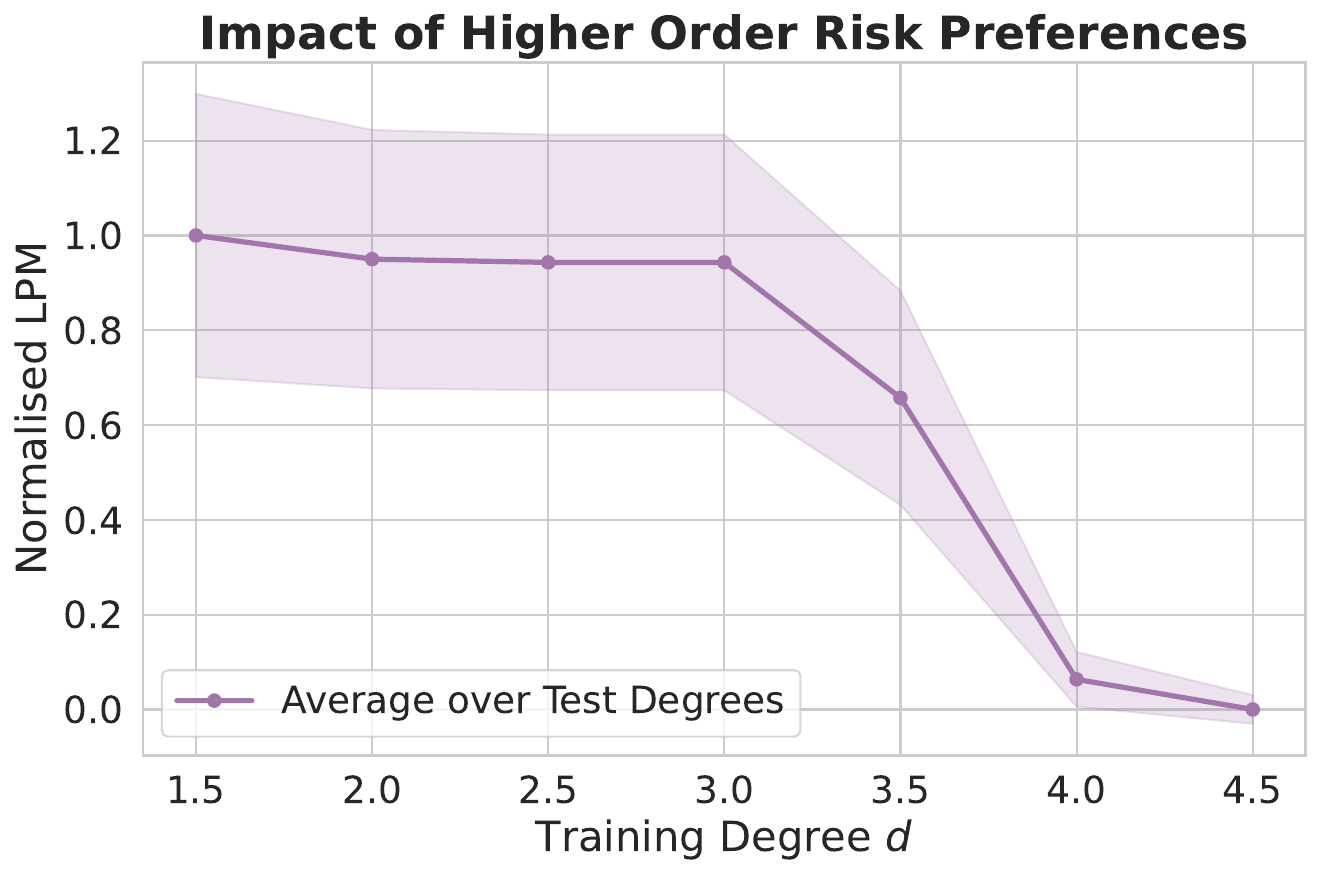}
    \caption{The impact of higher-order preferences $d \in \{1.5,2,2.5,3,3.5,4\}$ on the mean risk (across all $d$). }
    %The  x-axis represents the trained degree, and the y-axis is the mean performance across the degrees. 
   %This performance is the test performance of that trained equilibrium on risk matrices of the other degrees (same underlying payoff matrix $\boldsymbol{M}$). 
   
    %For example, $x=4.5$ represents $\boldsymbol{\sigma}^{4.5}$, trained to optimise the utility when using $\Sigma^{LPM}(d=4.5)$, and evaluated across the various risk matrices $\Sigma^{LPM}(\{1.5,2,2.5,3,3.5,4\})$.
    \label{figHigherOrder}
    \end{minipage}\hfill
\end{figure*}

\textit{Threshold $\boldsymbol{\tau}$} The combination of the risk aversion $\gamma$ and the LPM threshold $\tau$ balances the expected reward achievable in DRAE, altering the resulting equilibria. \cref{figLowTau} demonstrates how the expected return of the DRAE solution changes over different $\gamma$ and $\tau$ values. For higher threshold values (which leads to more strategies having high downside risk), as $\gamma$ increases the expected return decreases more rapidly as DRAE prioritises reducing the downside risk over expected return.  As we increase $\tau$ and/or $\gamma$, more conservative (w.r.t. downside risk) strategies are selected, with the most conservative strategies with the highest threshold $\tau=100$ and $\gamma$, at the expense of the expected return. %The threshold value of $\tau=0$ (no strategies produce any downside risk) maximises the expected return for all levels of risk-aversion $\gamma$. 

\subsection{Asset Game}
\paragraph{Environment}
The asset market game \cite{alos2005asset}, models a strategic market with a fixed supply of assets and outside fiat money, as a static game played only once by strategic investors. Players (investors) have wealth $w_1 = w_2 = 0.5$, which they can allocate over $M$ assets $as_1, ..., as_M$ by choosing portfolios (their actions). A portfolio for player $i$ is an allocation of their wealth over the assets: 
$
P_i = [p_i^{as_1}, \dots, p_i^{as_M}]
$
where $\sum_{m=1}^M p_i^{as_m} = 1$.
%actions involve allocating wealth across available assets. The risk profiles of the assets depends on which state of the environment is drawn from a state probability distribution. In addition, investors face the risk of under-diversifying their portfolios and sharing similar portfolios to the other investors, significantly reducing their overall returns as the proportion of asset returns is shared. 
Investor decisions are driven by the returns of their portfolio, and therefore the returns of the individual assets themselves. The returns of the individual assets are state-dependent, allowing for exogenous risk. At the start of the game, a state $s \in \mathcal{S}$ is drawn randomly, and the returns of assets depend on this state. Each asset has a set return in each state, $r^{as_m}(s)$, where these returns are set in the same manner as that defined for the synthetic games in \cref{sec:synthetic}. In other words, some assets are highly skewed and some have high variance which is consistent with real-world findings \cite{pastor2012stocks, albuquerque2012skewness}.

Returns of an asset are different to the final reward to an investor. The payoffs to the investor also depend on the amount of demand for an asset and therefore the corresponding price which they can acquire the asset at. In this setting, asset prices are defined by the amount of overall investor wealth that is put towards an asset. Consider the $m$-th asset $as_m$, the price of this asset $as_m$ would be $pr^{as_m} = \sum_{i=1}^2 p_i^{as_m} \cdot w_i$, which measures the total amount of wealth invested in an asset. Based on the prices, and the amount of wealth allocated by each investor, amounts of the asset are then distributed between the investors. This distribution is proportional to the investor allocation in the asset, if they allocated more then they will have more of the asset distributed to them:
$
    all^{as_m}_i = \frac{p_i^{as_m} \cdot w_i}{pr^{as_m}}
$
The payoff that an investor receives is:

\begin{align}
    U_i(P_i) = \sum_{m=1}^M \sum_{s \in \mathcal{S}} q(s) \cdot \Big[r^{as_m}(s) \cdot all_i^{as_m}\Big]
\end{align}
where $q(s)$ is the probability of state $s$ being drawn. Intuitively, $U_i$ rises by finding portfolios with larger returns over states, where the contained assets are underinvested.%by picking assets that the other investor has not selected in their portfolio. 

%In order to create a NFG, we cannot simply treat each $as_m$ as a NFG action. This is not appropriate as constructing a portfolio by mixing over these actions is not equivalent to an actual portfolio. Consider a 2 asset scenario and let $P^1 = [1, 0], P^2=[0,1]$ and $P^3 = [0.5, 0.5]$. We cannot form an NFG out of $P^1$ and $P^2$ as $(0.5P^1 + 0.5P^2) \neq P^3$. Therefore, we instead generate a selection of portfolios than an investor is able to pick from. All of our experiments use 100 generated portfolios. Out of these portfolios, we generated 80 that followed a Sobol sequence to guarantee diversity amongst portfolios. One problem with these is that in all portfolios, every asset receives non-zero investment proportion. Therefore, for our final 20 portfolios we randomly sample an amount of zero-proportion assets from $1$ to $M$ and then randomly generate portfolios assuming they have a certain amount of zero asset proportions.

In our experiments, we create 100 sample portfolios over 10 assets (using Sobol sequencing). There are 5 states with probabilities $q(s) \sim \mathcal{U}[0,1]$ which are normalised such that $\sum_{s \in \mathcal{S}}q(s) = 1$. The threshold $\tau$ is set to be the expected return if we randomly selected one of the portfolios. 

\paragraph{Results}

The results are shown in \cref{figAssetVariance} and \cref{figAssetDownside}. Similar overall trends to the synthetic game holds, demonstrating that DRAE can be successfully applied to these more complex state dependent environments. While in the synthetic game we saw that DRAE minimised variance, in \cref{figAssetVariance} we see that DRAE does not meaningfully reduce the overall variance of its solutions. However, as in the synthetic game, DRAE is able to greatly decrease the downside risk of its solutions. These results suggests that, in the process of minimising downside risk, the potential for upside returns increases (as the total variance is staying the same). Under the financial setting described here, this is particularly desirable behaviour as it means not only do the selected portfolios have lower downside risk, but they allow for higher potential upside profit as well. This behaviour demonstrates one of the key properties of DRAE: \emph{the potential for upside risk is unconstrained.} 
%Whilst the optimal goal would be to increase the expected reward as much as possible, a near second goal would be to allow for unconstrained upside risk - i.e. returns far above the threshold

In contrast, the RAE results highlight again the misleading performance of minimising variance as a measure of risk. While the RAE solutions have very low total variance, as the risk-aversion increases, the downside risk of these RAE solutions actually increases (\cref{figAssetDownside}). This suggests that RAE finds portfolio mixtures with a consistent range of returns, however, the returns for these portfolios generally fall below the threshold $\tau$, as indicated by the positive LPM in \cref{figAssetDownside}. This poor performance is because the return distribution has less variation in this lower payoff region, causing RAE to narrow in to these portfolios, and as a result, missing out on many of the portfolios which have higher potential upside (returns$ > \tau$). DRAE remedies this issue by minimising only on the downside, not restricting variation above $\tau$.

\textit{States $\mathcal{S}$} We compare the performance (on downside risk) of DRAE and RAE as we increase the amount of states in the environment (\cref{figEndogenousRisk}). Increasing the amount of states leads to more risk as the returns for portfolios become more stochastic. As the number of states increases, the LPM
for RAE (slightly) increases, suggesting that RAE is not fully accounting for the additional exogenous risk. On the other hand, DRAE successfully handles this exogenous risk,
with the differences between the two approaches becoming more pronounced as the number of states increases.
%demonstrates a large reduction in downside risk upon the introduction of states, with largest drop happening from 1 states to 2 states. 
%suggesting that DRAE reacts accordingly to the increase in risk introduced by the states by finding more risk-averse equilibria.
%as this distribution narrows, many value of the return sits beneath the pre-defined threshold $\tau$ - the mean returns from the uniform strategy. If we consider $\tau$ to represent a minimum amount of return a player requires to willingly partake in the asset market, the RAE approach may be low variance but consistently performs below this desired level of return. 

\subsection{Product Portfolio Game}
\paragraph{Environment}
The Product Portfolio Management game (PPM) \cite{sadeghi2011game} models the general economic problem of understanding production ability, market potential, and how these combine to determine the best set of products to offer. PPM models a competitive environment between firms aiming to capture market demand. The risk comes from both the stochasticity of the market demand and product competition.% the risk that the competitor directly enters a market you are heavily invested in, suggesting under-diversification across markets as a potential cause of risky strategies. %The PPM game has potential applications in the financial setting from mortgage product to credit card product offerings.

In PPM, each player is able to develop $M$ products $p_1,...,p_M$. Each product has an associated cost of development $c_1, ..., c_M$. There are $L$ market segments $K = \{k_1,...,k_L\}$ in which the products receive different utilities, where $u_{mk}$ represents the utility of product $m$ in segment $k$. For example, different market segments could represent different age groups, where a product may drive higher demand with e.g. older consumers. Players are able to design \textit{portfolios} $P$ over these ($2^M - 1$) potential products combinations.

The payoff to the player $i$ of a product portfolio, $P_i^\prime$ depends on the segment utilities, the product costs, and the market competition. 
%Additionally, the payoff depends on the product portfolio of the competing player $P_j^\prime$. This represents that the competing products put pressure on the overall demand of a market segment. There are two main factors to consider when setting the payoff of a product: 1) the utility to the player with respect to the cost of the product and 2) the total market share captured. 
The payoff function is based on the customer-engineering interaction model of \cite{jiao2005product}:

\begin{align}
E = \sum_{k \in K} \sum_{p=1}^{|P_i^\prime|} \left( \frac{u_{pk}}{c_p} \cdot D_{pk} \cdot Q_k \right)
\end{align}
representing the expected surplus for player $i$ of their product portfolio $P_i^\prime$. $Q_k$ represents the size of a market segment $k$. The demand $D_{pk}$ is based on the market size $Q_k$, competing products $P_{com} = \{p' \in P_i^\prime, P_j^\prime\}$, and the (relative) utility of the product $u_{pk}$:
$
    D_{pk} = \frac{e^{\mu \cdot u_{pk}}}{\sum_{m \in P_{com}} e^{\mu \cdot u_{mk}}} \cdot Q_k
$
\footnote{$\mu$ is just a scaling parameter controlling utility impact}%, and $P_{com} = \{P_i^\prime, P_j^\prime\}$ is set of all of the products in the market including those from the players own portfolio and those of the competing portfolio.
%Furthermore, the fewer competing products $N_{com}$ also the larger the choice probability.
This allows us to define the payoff function $f$ to player $i$ of using portfolio $P_i^\prime$ when player $j$ selects portfolio $P_j^\prime$:
\begin{align}
    \boldsymbol{M}_{i,j} = \sum_{k \in K} \sum_{p=1}^{|P_i^\prime|} \Big( \frac{u_{pk}}{c_p} \cdot \frac{e^{\mu} \cdot u_{pk}}{\sum_{m=1}^{N_{com}} e^{\mu \cdot u_{mk}}} \cdot Q_k\Big)
\end{align}

\paragraph{Results}
The results are shown in \cref{figPPMVar,figPPMDown}. Similar, but more extreme, trends to the other two environments hold, demonstrating the robustness in performance of DRAE across the various environments. DRAE monotonically decrease both variance and downside risk as $\gamma$ increases. RAE, however, demonstrates far more extreme changes over $\gamma$ - for example while variance reduces substantially, the increase in downside risk is near unbounded, resulting in product strategies with high downside risk at the equilibrium. The axes of \cref{figPPMVar} and \cref{figPPMDown} are intentionally clipped so that easier comparisons between DRAE and RAE can be made due to the scale. For example, for the highest $\gamma$ tested on, the LPM value reaches 21396 for RAE even though the variance has been substantially reduced.

\textit{Degree ${d}$} The impact of the degree $d$ on the resulting downside risk is visualised in \cref{figHigherOrder}. As we increase the degree which DRAE optimises for, the resulting risk reduces, as the players become more focused on these higher order preferences which generally capture a lot of the lower-order risks implicitly in their solution. While the degree is often problem specific, and may vary in different domains \cite{georgalos2023higher}, the goal of \cref{figHigherOrder} is just to demonstrate that DRAE is flexible enough to encode any of these risk preferences. Setting a degree which is too low can lead to the model missing higher-order parts of the risk distribution, which may be important to the players.

\subsection{Key Takeaways}
When designing DRAE we had three desiderata: 1) to optimise for downside variance only, 2) flexibility in terms of the risk preferences, and 3) to consider exogenous risk as well as risk caused by the opposing players. In our experimental results we demonstrated that DRAE met all of these criteria: 1) it significantly outperforms mean-variance solution concepts in terms of reducing the downside variance (\cref{figResults}) over all of our environments, 2) we can alter the degree of $\Sigma^{LPM}$ to capture differing orders of risk with higher-orders still capturing lower-order preferences (\cref{figHigherOrder}) and the threshold value $\tau$ (\cref{figLowTau}), and 3) we demonstrated that DRAE is able to handle the impact of exogenous risk where RAE fails to (\cref{figEndogenousRisk}). 

%\section{Discussion}

\section{Conclusions}
In this work, we proposed DRAE, a risk-aware game-theoretic equilibrium concept that utilises LPMs as a risk measure that remedies the negative properties of existing reward-variance approaches. DRAE improves over existing variance-aware RAE on multiple fronts: 1) by accounting for exclusively downside risk created by the strategy of the opposing players, 2) by introducing flexibility to model higher-order risk attitudes and 3) by modelling the risk inherent to the environment (e.g. asset price fluctations). DRAE is fundamentally designed to be an equilibrium tool for risky environments, where there exists stochasticity in the environment and little incentive to constrain potential upside in the face of this stochasticity (e.g., in finance). We demonstrated the applicability of the proposed approach on several example environments, consistently outperforming the existing equilibrium approaches in all environments by limiting the downside risk of the selected strategies. Additionally, we proved the existence of this DRAE, and guarantees of minimising downside risk for a given level of expected return. Future work should focus on computing DRAE in temporal environments, for example through the use of reinforcement learning (RL). Game-theoretic solution frameworks, such as policy-space response oracles \cite{lanctot2017unified}, would require an RL algorithm that could optimise for both the expected cumulative reward and LPM of the cumulative reward of a temporal setting.

\small{
\subsection*{}
\textbf{Disclaimer} This paper was prepared for informational purposes in part by the Artificial Intelligence Research group of JPMorgan Chase \& Co. and its affiliates (``JP Morgan'') and is not a product of the Research Department of JP Morgan. JP Morgan makes no representation and warranty whatsoever and disclaims all liability, for the completeness, accuracy or reliability of the information contained herein. This document is not intended as investment research or investment advice, or a recommendation, offer or solicitation for the purchase or sale of any security, financial instrument, financial product or service, or to be used in any way for evaluating the merits of participating in any transaction, and shall not constitute a solicitation under any jurisdiction or to any person, if such solicitation under such jurisdiction or to such person would be unlawful.
}

\newpage

\bibliographystyle{elsarticle-num-names}
\bibliography{bib}

\begin{thebibliography}{35}
\expandafter\ifx\csname natexlab\endcsname\relax\def\natexlab#1{#1}\fi
\providecommand{\url}[1]{\texttt{#1}}
\providecommand{\href}[2]{#2}
\providecommand{\path}[1]{#1}
\providecommand{\DOIprefix}{doi:}
\providecommand{\ArXivprefix}{arXiv:}
\providecommand{\URLprefix}{URL: }
\providecommand{\Pubmedprefix}{pmid:}
\providecommand{\doi}[1]{\href{http://dx.doi.org/#1}{\path{#1}}}
\providecommand{\Pubmed}[1]{\href{pmid:#1}{\path{#1}}}
\providecommand{\bibinfo}[2]{#2}
\ifx\xfnm\relax \def\xfnm[#1]{\unskip,\space#1}\fi
%Type = Incollection
\bibitem[{Bielefeld(1988)}]{bielefeld1988reexamination}
\bibinfo{author}{R.~S. Bielefeld},
\newblock \bibinfo{title}{Reexamination of the perfectness concept for equilibrium points in extensive games},
\newblock in: \bibinfo{booktitle}{Models of Strategic Rationality}, \bibinfo{publisher}{Springer}, \bibinfo{year}{1988}, pp. \bibinfo{pages}{1--31}.
%Type = Article
\bibitem[{McKelvey and Palfrey(1995)}]{mckelvey1995quantal}
\bibinfo{author}{R.~D. McKelvey}, \bibinfo{author}{T.~R. Palfrey},
\newblock \bibinfo{title}{Quantal response equilibria for normal form games},
\newblock \bibinfo{journal}{Games and economic behavior} \bibinfo{volume}{10} (\bibinfo{year}{1995}) \bibinfo{pages}{6--38}.
%Type = Inproceedings
\bibitem[{Slumbers et~al.(2023)Slumbers, Mguni, Blumberg, Mcaleer, Yang, and Wang}]{slumbers2023game}
\bibinfo{author}{O.~Slumbers}, \bibinfo{author}{D.~H. Mguni}, \bibinfo{author}{S.~B. Blumberg}, \bibinfo{author}{S.~M. Mcaleer}, \bibinfo{author}{Y.~Yang}, \bibinfo{author}{J.~Wang},
\newblock \bibinfo{title}{A game-theoretic framework for managing risk in multi-agent systems},
\newblock in: \bibinfo{booktitle}{International Conference on Machine Learning}, \bibinfo{organization}{PMLR}, \bibinfo{year}{2023}, pp. \bibinfo{pages}{32059--32087}.
%Type = Techreport
\bibitem[{Egan et~al.(2021)Egan, MacKay, and Yang}]{egan2021drives}
\bibinfo{author}{M.~L. Egan}, \bibinfo{author}{A.~MacKay}, \bibinfo{author}{H.~Yang}, \bibinfo{title}{What drives variation in investor portfolios? estimating the roles of beliefs and risk preferences}, \bibinfo{type}{Technical Report}, National Bureau of Economic Research, \bibinfo{year}{2021}.
%Type = Article
\bibitem[{Nawrocki(1991)}]{nawrocki1991optimal}
\bibinfo{author}{D.~N. Nawrocki},
\newblock \bibinfo{title}{Optimal algorithms and lower partial moment: ex post results},
\newblock \bibinfo{journal}{Applied Economics} \bibinfo{volume}{23} (\bibinfo{year}{1991}) \bibinfo{pages}{465--470}.
%Type = Article
\bibitem[{Harsanyi et~al.(1988)Harsanyi, Selten et~al.}]{harsanyi1988general}
\bibinfo{author}{J.~C. Harsanyi}, \bibinfo{author}{R.~Selten}, et~al.,
\newblock \bibinfo{title}{A general theory of equilibrium selection in games},
\newblock \bibinfo{journal}{MIT Press Books} \bibinfo{volume}{1} (\bibinfo{year}{1988}).
%Type = Article
\bibitem[{Nash(1951)}]{nash1951non}
\bibinfo{author}{J.~Nash},
\newblock \bibinfo{title}{Non-cooperative games},
\newblock \bibinfo{journal}{Annals of mathematics}  (\bibinfo{year}{1951}) \bibinfo{pages}{286--295}.
%Type = Inproceedings
\bibitem[{Mazumdar et~al.(2024)Mazumdar, Panaganti, and Shi}]{mazumdar2024a}
\bibinfo{author}{E.~Mazumdar}, \bibinfo{author}{K.~Panaganti}, \bibinfo{author}{L.~Shi},
\newblock \bibinfo{title}{A behavioral economics approach to principled multi-agent reinforcement learning},
\newblock in: \bibinfo{booktitle}{NeurIPS 2024 Workshop on Behavioral Machine Learning}, \bibinfo{year}{2024}. \URLprefix \url{https://openreview.net/forum?id=SKWfQtfR3v}.
%Type = Article
\bibitem[{Royset(2022)}]{royset2022risk}
\bibinfo{author}{J.~O. Royset},
\newblock \bibinfo{title}{Risk-adaptive approaches to learning and decision making: A survey},
\newblock \bibinfo{journal}{arXiv preprint arXiv:2212.00856}  (\bibinfo{year}{2022}).
%Type = Article
\bibitem[{Yekkehkhany et~al.(2020)Yekkehkhany, Murray, and Nagi}]{yekkehkhany2020risk}
\bibinfo{author}{A.~Yekkehkhany}, \bibinfo{author}{T.~Murray}, \bibinfo{author}{R.~Nagi},
\newblock \bibinfo{title}{Risk-averse equilibrium for games},
\newblock \bibinfo{journal}{arXiv preprint arXiv:2002.08414}  (\bibinfo{year}{2020}).
%Type = Inproceedings
\bibitem[{Xu(2024)}]{xu2024balancing}
\bibinfo{author}{Y.~Xu},
\newblock \bibinfo{title}{Balancing risk and reward: Cfrra for extensive-form games},
\newblock in: \bibinfo{booktitle}{Fourth International Conference on Advanced Algorithms and Neural Networks (AANN 2024)}, volume \bibinfo{volume}{13416}, \bibinfo{organization}{SPIE}, \bibinfo{year}{2024}, pp. \bibinfo{pages}{647--653}.
%Type = Article
\bibitem[{Markowitz(1952)}]{10.2307/2975974}
\bibinfo{author}{H.~Markowitz},
\newblock \bibinfo{title}{Portfolio selection},
\newblock \bibinfo{journal}{The Journal of Finance} \bibinfo{volume}{7} (\bibinfo{year}{1952}) \bibinfo{pages}{77--91}. \URLprefix \url{http://www.jstor.org/stable/2975974}.
%Type = Article
\bibitem[{Estrada(2004)}]{estrada2004mean}
\bibinfo{author}{J.~Estrada},
\newblock \bibinfo{title}{Mean-semivariance behaviour: an alternative behavioural model},
\newblock \bibinfo{journal}{Journal of Emerging Market Finance} \bibinfo{volume}{3} (\bibinfo{year}{2004}) \bibinfo{pages}{231--248}.
%Type = Article
\bibitem[{Ferrari et~al.(2024)Ferrari, Paterlini, Rigamonti, and Weissensteiner}]{ferrari2024smoothed}
\bibinfo{author}{D.~Ferrari}, \bibinfo{author}{S.~Paterlini}, \bibinfo{author}{A.~Rigamonti}, \bibinfo{author}{A.~Weissensteiner},
\newblock \bibinfo{title}{Smoothed semicovariance estimation for portfolio selection},
\newblock \bibinfo{journal}{Available at SSRN 4789945}  (\bibinfo{year}{2024}).
%Type = Article
\bibitem[{Markowitz et~al.(2020)Markowitz, Starer, Fram, and Gerber}]{markowitz2020avoiding}
\bibinfo{author}{H.~M. Markowitz}, \bibinfo{author}{D.~Starer}, \bibinfo{author}{H.~Fram}, \bibinfo{author}{S.~Gerber},
\newblock \bibinfo{title}{Avoiding the downside: A practical review of the critical line algorithm for mean--semivariance portfolio optimization},
\newblock \bibinfo{journal}{Handbook of applied investment research}  (\bibinfo{year}{2020}) \bibinfo{pages}{369--415}.
%Type = Article
\bibitem[{{\~N}{\'\i}guez et~al.(2015){\~N}{\'\i}guez, Paya, Peel, and Perote}]{niguez2015higher}
\bibinfo{author}{T.~M. {\~N}{\'\i}guez}, \bibinfo{author}{I.~Paya}, \bibinfo{author}{D.~A. Peel}, \bibinfo{author}{J.~Perote},
\newblock \bibinfo{title}{Higher-order risk preferences, constant relative risk aversion and the optimal portfolio allocation}  (\bibinfo{year}{2015}).
%Type = Article
\bibitem[{Mondal and Selvaraju(2022)}]{mondal2022convexity}
\bibinfo{author}{D.~Mondal}, \bibinfo{author}{N.~Selvaraju},
\newblock \bibinfo{title}{Convexity, two-fund separation and asset ranking in a mean-lpm portfolio selection framework},
\newblock \bibinfo{journal}{OR Spectrum} \bibinfo{volume}{44} (\bibinfo{year}{2022}) \bibinfo{pages}{225--248}.
%Type = Article
\bibitem[{Nawrocki(1992)}]{nawrocki1992characteristics}
\bibinfo{author}{D.~N. Nawrocki},
\newblock \bibinfo{title}{The characteristics of portfolios selected by n-degree lower partial moment},
\newblock \bibinfo{journal}{International Review of Financial Analysis} \bibinfo{volume}{1} (\bibinfo{year}{1992}) \bibinfo{pages}{195--209}.
%Type = Article
\bibitem[{Cumova and Nawrocki(2011)}]{cumova2011symmetric}
\bibinfo{author}{D.~Cumova}, \bibinfo{author}{D.~Nawrocki},
\newblock \bibinfo{title}{A symmetric lpm model for heuristic mean--semivariance analysis},
\newblock \bibinfo{journal}{Journal of Economics and Business} \bibinfo{volume}{63} (\bibinfo{year}{2011}) \bibinfo{pages}{217--236}.
%Type = Article
\bibitem[{Estrada(2008)}]{estrada2008mean}
\bibinfo{author}{J.~Estrada},
\newblock \bibinfo{title}{Mean-semivariance optimization: A heuristic approach},
\newblock \bibinfo{journal}{Journal of Applied Finance (Formerly Financial Practice and Education)} \bibinfo{volume}{18} (\bibinfo{year}{2008}).
%Type = Article
\bibitem[{Higham(1988)}]{higham1988computing}
\bibinfo{author}{N.~J. Higham},
\newblock \bibinfo{title}{Computing a nearest symmetric positive semidefinite matrix},
\newblock \bibinfo{journal}{Linear algebra and its applications} \bibinfo{volume}{103} (\bibinfo{year}{1988}) \bibinfo{pages}{103--118}.
%Type = Article
\bibitem[{Merton(1972)}]{merton1972analytic}
\bibinfo{author}{R.~C. Merton},
\newblock \bibinfo{title}{An analytic derivation of the efficient portfolio frontier},
\newblock \bibinfo{journal}{Journal of financial and quantitative analysis} \bibinfo{volume}{7} (\bibinfo{year}{1972}) \bibinfo{pages}{1851--1872}.
%Type = Article
\bibitem[{Fudenberg and Kreps(1993)}]{fudenberg1993learning}
\bibinfo{author}{D.~Fudenberg}, \bibinfo{author}{D.~M. Kreps},
\newblock \bibinfo{title}{Learning mixed equilibria},
\newblock \bibinfo{journal}{Games and economic behavior} \bibinfo{volume}{5} (\bibinfo{year}{1993}) \bibinfo{pages}{320--367}.
%Type = Article
\bibitem[{Al{\'o}s-Ferrer and Ania(2005)}]{alos2005asset}
\bibinfo{author}{C.~Al{\'o}s-Ferrer}, \bibinfo{author}{A.~B. Ania},
\newblock \bibinfo{title}{The asset market game},
\newblock \bibinfo{journal}{Journal of mathematical economics} \bibinfo{volume}{41} (\bibinfo{year}{2005}) \bibinfo{pages}{67--90}.
%Type = Article
\bibitem[{Sadeghi and Zandieh(2011)}]{sadeghi2011game}
\bibinfo{author}{A.~Sadeghi}, \bibinfo{author}{M.~Zandieh},
\newblock \bibinfo{title}{A game theory-based model for product portfolio management in a competitive market},
\newblock \bibinfo{journal}{Expert Systems with Applications} \bibinfo{volume}{38} (\bibinfo{year}{2011}) \bibinfo{pages}{7919--7923}.
%Type = Article
\bibitem[{Azzalini and Valle(1996)}]{azzalini1996multivariate}
\bibinfo{author}{A.~Azzalini}, \bibinfo{author}{A.~D. Valle},
\newblock \bibinfo{title}{The multivariate skew-normal distribution},
\newblock \bibinfo{journal}{Biometrika} \bibinfo{volume}{83} (\bibinfo{year}{1996}) \bibinfo{pages}{715--726}.
%Type = Article
\bibitem[{P{\'a}stor and Stambaugh(2012)}]{pastor2012stocks}
\bibinfo{author}{L.~P{\'a}stor}, \bibinfo{author}{R.~F. Stambaugh},
\newblock \bibinfo{title}{Are stocks really less volatile in the long run?},
\newblock \bibinfo{journal}{The Journal of Finance} \bibinfo{volume}{67} (\bibinfo{year}{2012}) \bibinfo{pages}{431--478}.
%Type = Article
\bibitem[{Albuquerque(2012)}]{albuquerque2012skewness}
\bibinfo{author}{R.~Albuquerque},
\newblock \bibinfo{title}{Skewness in stock returns: Reconciling the evidence on firm versus aggregate returns},
\newblock \bibinfo{journal}{The Review of Financial Studies} \bibinfo{volume}{25} (\bibinfo{year}{2012}) \bibinfo{pages}{1630--1673}.
%Type = Article
\bibitem[{Jiao and Zhang(2005)}]{jiao2005product}
\bibinfo{author}{J.~Jiao}, \bibinfo{author}{Y.~Zhang},
\newblock \bibinfo{title}{Product portfolio planning with customer-engineering interaction},
\newblock \bibinfo{journal}{Iie Transactions} \bibinfo{volume}{37} (\bibinfo{year}{2005}) \bibinfo{pages}{801--814}.
%Type = Article
\bibitem[{Georgalos et~al.(2023)Georgalos, Paya, and Peel}]{georgalos2023higher}
\bibinfo{author}{K.~Georgalos}, \bibinfo{author}{I.~Paya}, \bibinfo{author}{D.~Peel},
\newblock \bibinfo{title}{Higher order risk attitudes: new model insights and heterogeneity of preferences},
\newblock \bibinfo{journal}{Experimental Economics} \bibinfo{volume}{26} (\bibinfo{year}{2023}) \bibinfo{pages}{145--192}.
%Type = Inproceedings
\bibitem[{Lanctot et~al.(2017)Lanctot, Zambaldi, Gruslys, Lazaridou, Tuyls, P{\'e}rolat, Silver, and Graepel}]{lanctot2017unified}
\bibinfo{author}{M.~Lanctot}, \bibinfo{author}{V.~Zambaldi}, \bibinfo{author}{A.~Gruslys}, \bibinfo{author}{A.~Lazaridou}, \bibinfo{author}{K.~Tuyls}, \bibinfo{author}{J.~P{\'e}rolat}, \bibinfo{author}{D.~Silver}, \bibinfo{author}{T.~Graepel},
\newblock \bibinfo{title}{A unified game-theoretic approach to multiagent reinforcement learning},
\newblock in: \bibinfo{booktitle}{Proceedings of the 31st International Conference on Neural Information Processing Systems}, \bibinfo{year}{2017}, pp. \bibinfo{pages}{4193--4206}.
%Type = Article
\bibitem[{Monderer and Shapley(1996)}]{monderer1996fictitious}
\bibinfo{author}{D.~Monderer}, \bibinfo{author}{L.~S. Shapley},
\newblock \bibinfo{title}{Fictitious play property for games with identical interests},
\newblock \bibinfo{journal}{Journal of economic theory} \bibinfo{volume}{68} (\bibinfo{year}{1996}) \bibinfo{pages}{258--265}.
%Type = Article
\bibitem[{Robinson(1951)}]{10.2307/1969530}
\bibinfo{author}{J.~Robinson},
\newblock \bibinfo{title}{An iterative method of solving a game},
\newblock \bibinfo{journal}{Annals of Mathematics} \bibinfo{volume}{54} (\bibinfo{year}{1951}) \bibinfo{pages}{296--301}. \URLprefix \url{http://www.jstor.org/stable/1969530}.
%Type = Article
\bibitem[{Goldberg et~al.(2013)Goldberg, Savani, S{\o}rensen, and Ventre}]{goldberg2013approximation}
\bibinfo{author}{P.~W. Goldberg}, \bibinfo{author}{R.~Savani}, \bibinfo{author}{T.~B. S{\o}rensen}, \bibinfo{author}{C.~Ventre},
\newblock \bibinfo{title}{On the approximation performance of fictitious play in finite games},
\newblock \bibinfo{journal}{International Journal of Game Theory} \bibinfo{volume}{42} (\bibinfo{year}{2013}) \bibinfo{pages}{1059--1083}.
%Type = Article
\bibitem[{Ganzfried(2020)}]{ganzfried2020fictitious}
\bibinfo{author}{S.~Ganzfried},
\newblock \bibinfo{title}{Fictitious play outperforms counterfactual regret minimization},
\newblock \bibinfo{journal}{arXiv preprint arXiv:2001.11165}  (\bibinfo{year}{2020}).

\end{thebibliography}

\clearpage
\appendix

\section{SFP Convergence}\label{appendixSFP}

SFP has convergence guarantees in a selection of games, e.g. potential games \cite{monderer1996fictitious} and finite two-player zero-sum games \cite{10.2307/1969530}, and
is known to be robust empirically in game classes not listed \cite{goldberg2013approximation, ganzfried2020fictitious, slumbers2023game}. In this section, we prove this convergence.

For a perturbed utility function $\bar{u}^i$ to be permissible in an SFP process, there exist two conditions:

\begin{enumerate}
    \item That there exists a unique global solution to $\bar{u}^i$.
    \item That the $\text{argmax}$ assigns strictly positive probability to all pure strategies.
\end{enumerate}
\noindent replacing the best-response \cref{eqBR} with the best-response map \cref{eq:br-map} satisfies the conditions of $\bar{u}^i$ for a SFP process, as shown below.

\begin{proof}

We let $\bar{u}^i$ be replaced by the best-response map \cref{eq:br-map} and show that the 2 conditions noted above are met.

\textbf{Condition 1 -} To show that there exists a global solution to $\bar{u}^i$ we need to show that $\bar{u}^i$ is strictly concave which guarantees a unique global maximum.  As the ER term $\boldsymbol{\sigma}^T \cdot \boldsymbol{M} \cdot \boldsymbol{\varsigma}$ is linear, we therefore require that the perturbation term $v^i(\boldsymbol{\sigma})$ is strictly convex such that the perturbed utility function $\bar{u}^i = \text{argmax}_{\boldsymbol{\sigma}} \operatorname{ER} - \lambda v^i(\boldsymbol{\sigma})$ is strictly concave. In \cref{appendixExistence} we have already shown that, as long as the risk matrix $\boldsymbol{\Sigma}^{\text{LPM}}$ is positive definite, then the quadratic term $\boldsymbol{\sigma}^{T} \cdot \boldsymbol{\Sigma}^{\text{LPM}} \cdot \boldsymbol{\sigma}$ is strictly convex. By design, $\boldsymbol{\Sigma}^{\text{LPM}}$ is strictly convex, thus $\bar{u}^i$ is also strictly concave and therefore has a unique global maximum.

\textbf{Condition 2 -}
By design the QP that solves \cref{eq:br-map} is constrained such that all pure strategies receive strictly positive probability (as is done in THPE and others), ensuring exploration of all strategies due to the constraint $\sigma(a) \geq \epsilon, \epsilon > 0$. Therefore, the argmax also assigns positive probabilities, and condition 2 is satisfied.

As our formulation meets these two conditions, SFP can be used to find DRAE in two-player zero-sum games and potential games.
\end{proof}

\end{document}